\documentclass[twocolumn]{IEEEtran}

\usepackage{cite}
\usepackage{amsfonts,amsmath,amssymb}
\usepackage{amsthm}
\usepackage{graphicx,color,epsfig,rotating}
\usepackage{latexsym}
\usepackage{subfigure}
\usepackage{enumitem}
\usepackage{cite}
\usepackage{epic,eepic}
\usepackage{algorithm}
\usepackage{algpseudocode}
\usepackage{algpseudocode}
\usepackage{float}
\usepackage{multirow}
\usepackage{booktabs} 
\def\BibTeX{{\rm B\kern-.05em{\sc i\kern-.025em b}\kern-.08em   T\kern-.1667em\lower.7ex\hbox{E}\kern-.125emX}}

\usepackage[linktoc=page,colorlinks=true,linkcolor=black,citecolor=black]{hyperref}

\usepackage{bbm}

\usepackage{bm}

\usepackage{orcidlink}

\allowdisplaybreaks

\newtheorem{remark}{Remark}
\definecolor{caribbeangreen}{rgb}{0.0, 0.8, 0.6}

\begin{document}

\title{Covert Multi-Hop Communications for Heterogeneous Networks With Multiple Wardens}

\author{ \IEEEauthorblockN{Justin H. Kong\raisebox{0.5ex}{\orcidlink{0000-0003-2856-7060}},~\emph{Senior Member, IEEE}, Terrence J. Moore\raisebox{0.5ex}{\orcidlink{0000-0003-3279-2965}},~\emph{Member, IEEE}, and \\  Fikadu T. Dagefu\raisebox{0.5ex}{\orcidlink{0000-0002-7532-5278}},~\emph{Senior Member, IEEE}  } \\

\thanks{The authors are with the U.S. Army Combat Capabilities Development Command (DEVCOM) Army Research Laboratory, Adelphi, MD 20783, USA (e-mail: justin.h.kong2.civ@army.mil; terrence.j.moore.civ@army.mil; fikadu.t.dagefu.civ@army.mil).
} } \maketitle



\begin{abstract}
This paper investigates covert multi-hop communications in heterogeneous wireless networks monitored by multiple passive wardens.
To maximize network-wide covertness while satisfying a strict end-to-end rate requirement, we jointly optimize routing, modality selection, and transmit power.
Under a simultaneous multi-hop transmission scheme, we analyze the detection capabilities of two distinct warden models: colluding wardens employing a central fusion center, and non-colluding wardens operating independently.
For both models, we derive optimal detectors and exact expressions for the detection error probability (DEP).
In addition, to reduce the complexity of evaluating the DEP, we develop highly accurate closed-form approximations based on gamma moment matching and establish rigorous DEP lower bounds using Kullback-Leibler (KL) divergence.
Building on this theoretical foundation, we propose an efficient two-stage optimization algorithm that decouples link-level resource allocation from network-level path selection.
By translating the KL divergence bounds into a novel, low-complexity routing metric, which universally simplifies to a linear summation of signal-to-noise ratios, we substantially reduce the computational overhead compared to conventional per-hop detection-based metrics. 
Finally, numerical simulations validate the theoretical analysis and demonstrate the near-optimal performance of the proposed framework.

\end{abstract}


\IEEEpeerreviewmaketitle

\section{Introduction}

\IEEEPARstart{W}{hile}  conventional physical-layer security protects message content, covert communications aim to conceal the very existence of transmissions, a task that becomes particularly critical and complex in environments monitored by multiple distributed wardens~\cite{Yan:19}.
To reliably evade such widespread surveillance, intelligent multi-hop routing minimizes signal leakage through a sequence of low-power transmissions, while the exploitation of multiple heterogeneous communication modalities (i.e., wireless technologies with varying characteristics) provides crucial spectral degrees of freedom~\cite{Wu:18b}.
Motivated by these considerations, this paper focuses on the rigorous analysis and joint optimization of covert multi-hop communications in heterogeneous networks in the presence of multiple wardens.

\subsection{Related Work and Motivation}

Conventional physical-layer security techniques based on the wire-tap channel primarily aim to prevent adversaries from decoding confidential messages~\cite{Wyner:75}. 
However, in highly security-sensitive networks, the mere existence of a transmission can create a critical vulnerability, as potential adversaries (wardens) can exploit this emission to localize friendly assets and trigger hostile countermeasures. 
In this context, covert communication has been extensively studied as a mechanism to conceal the presence of the transmission, thereby enforcing a more stringent level of security~\cite{Chen:23}.

The fundamental information-theoretic limits of covert communications were established for various types of channel models including additive white Gaussian noise (AWGN)~\cite{Bash:13}, binary symmetric~\cite{Che:13}, discrete memoryless~\cite{Wang:16c}, and multiple-access ~\cite{Arumugam:16}. 
Leveraging these theoretical underpinnings, recent literature has investigated practical covert transmission strategies across diverse wireless environments, including intelligent reflecting surface (IRS)-enabled systems~\cite{Kong:21, Lv:22, Kong:24}, unmanned aerial vehicle (UAV) networks~\cite{Rao:22,Hou:23,Deng:24}, satellite communications (SATCOM)~\cite{Jia:25, Yu:25}, and integrated sensing and communications (ISAC) architectures~\cite{Ma:23,Wang:24}.

To achieve more stringent covertness requirements, the integration of multiple complementary communication modalities offering distinct center frequencies, as well as diverse, and complementary propagation characteristics has recently garnered significant interest.
For single-hop transmissions with single-modality selection, the enhancement of communication covertness has been studied for heterogeneous networks with low-VHF and microwave modalities~\cite{Kong:22WCNC} and UAV networks utilizing microwave and millimeter-wave (mmWave) bands~\cite{Zhang:24}.
Subsequent work advanced toward simultaneous multi-modal transmission strategies by considering the concurrent exploitation of two fixed modalities~\cite{Aggarwal:24} and generalizing this approach to simultaneous multi-modal selection across arbitrary bands~\cite{Kong:26arXiv}.

While the aforementioned multi-modal strategies significantly improve covertness for single-hop links, establishing reliable communication over extended distances necessitates multi-hop routing to strictly limit the transmit power required at any individual node.
In the context of multi-hop heterogeneous networks monitored by a single warden, foundational work initially established a centralized optimization framework for single-flow covert routing to maximize end-to-end covertness~\cite{Kong:24TIFS}.
To alleviate the overhead associated with centralized control, a decentralized single-flow routing scheme based on reinforcement learning (RL) was subsequently developed in~\cite{Kong:24CL}.
This decentralized RL framework was later extended to incorporate practical network challenges, including node failure resiliency~\cite{Kim:24} and threat region avoidance~\cite{Kim:25}.
Recognizing the inherent advantages of multi-modal architectures, recent research has begun integrating simultaneous multi-modal transmissions into the multi-hop paradigm.
For instance, optimal link configurations featuring simultaneous transmission and power control were analyzed for covert rate maximization in two-hop networks in~\cite{Gillani:25CISS}, while a decentralized energy-efficient covert routing protocol utilizing two fixed simultaneous modalities was proposed in~\cite{Haque:25}.
Finally, joint routing and resource allocation frameworks have very recently been expanded to support multi-flow covert communications~\cite{Kong:25}.

Despite the efficacy of these routing schemes against a single warden, practical security-sensitive environments are often monitored by multiple distributed wardens, significantly elevating the risk of detection and making the study of multi-warden evasion fundamentally important.
In the context of single-hop networks, early efforts to combat such surveillance utilized spatially distributed friendly jammers to generate artificial noise against multiple cooperative wardens~\cite{Soltani:18}.
Subsequent studies further explored single-hop covertness in the finite blocklength regime with randomly distributed colluding wardens~\cite{Ma:22}, as well as in environments featuring multiple interferers alongside colluding wardens~\cite{Zhao:25}.
To address the compounded vulnerability of long-distance communications, multi-hop routing strategies have also been explicitly investigated under the multi-warden threat.
Specifically, the foundational work in~\cite{Sheikholeslami:18} investigated rate-maximizing routing under a covertness constraint, utilizing a lower bound on the detection error probability (DEP) for single-modality networks.
More recently, this multi-warden routing paradigm was extended to heterogeneous architectures in~\cite{Gillani:25arXiv}, which developed optimal routing and link configurations based on DEP lower bounds for networks where multiple modalities are concurrently utilized on each link.

While existing multi-warden routing frameworks predominantly assume wardens process transmissions from different hops independently~\cite{Sheikholeslami:18,Gillani:25arXiv}, a worst-case scenario involves highly capable wardens exploiting the coherent superposition of simultaneous transmissions across the network.
Furthermore, prior covert routing studies in~\cite{Kong:24TIFS,Kong:24CL,Kim:24,Kim:25,Gillani:25CISS,Haque:25,Kong:25,Sheikholeslami:18,Gillani:25arXiv} rely on approximations or bounds on the DEP, and exact DEP analysis has not been investigated in the literature, even for the single-warden case.
Motivated by these gaps, this paper provides a rigorous theoretical analysis of the exact DEP and develops a low-complexity joint optimization framework for routing, modality selection, and transmit power in simultaneous multi-hop heterogeneous networks against both colluding and non-colluding wardens.

\subsection{Contributions and Organization}

In this paper, we analyze the DEP of covert multi-hop communications in heterogeneous networks monitored by multiple passive wardens and propose a low-complexity joint routing and resource optimization algorithm that maximizes the DEP subject to an end-to-end rate requirement.
Specifically, the main contributions are summarized as follows:
\begin{itemize}
    \item We formulate a worst-case adversarial scenario characterized by a simultaneous multi-hop transmission scheme, wherein wardens exploit the coherent superposition of aggregated signals. Under this framework, we identify the optimal detectors and exact expressions for the overall network DEP for two distinct threat models: colluding wardens utilizing a central fusion center and non-colluding wardens acting independently.
    
    \item To alleviate the computational intractability of evaluating the exact DEP, we derive highly accurate closed-form DEP approximations based on gamma moment matching of the test statistics. Furthermore, we establish analytically tractable lower bounds on the DEP using Kullback-Leibler (KL) divergence for both warden models. We also introduce alternative DEP formulations under a conventional detection model where wardens attempt to detect transmissions on a per-hop basis.
    
    \item Building upon the theoretical bounds, we propose an efficient two-stage optimization algorithm that decouples link-level resource allocation from network-level path selection to jointly optimize routing, modality selection, and transmit power. Specifically, by establishing that the optimal transmit power and modality for any candidate link are uniquely determined by a rate constraint, our framework transforms a highly coupled, mixed-integer problem into a scalable shortest-path search.
    
    \item We translate the KL divergence bounds into a novel low-complexity routing metric. Remarkably, we prove that this metric universally simplifies to a linear summation of signal-to-noise ratios (SNRs) for both colluding and non-colluding models. This formulation reduces the computational overhead compared to conventional per-hop detection proxies while seamlessly integrating with Dijkstra's algorithm. Lastly, through numerical simulations, we validate our analysis and demonstrate the efficiency of the proposed algorithm by comparing it to the optimal performance obtained via the exhaustive search method, which has an intractable combinatorial complexity.
    
\end{itemize}

The remainder of this paper is structured as follows.
Section~\ref{sec:network_model} introduces the network model, communication performance, and multi-warden detection models, followed by the problem formulation.
Sections~\ref{sec:colluding_analysis} and~\ref{sec:noncolluding_analysis} detail the exact DEP analysis, moment-matching approximations, KL divergence lower bounds, and alternative DEP formulations under a per-hop detection model for the colluding and non-colluding wardens, respectively.
In Section~\ref{sec:optimization}, we propose a novel low-complexity joint routing and resource allocation algorithm along with a comprehensive complexity analysis.
Section~\ref{sec:results} provides numerical simulations to validate the theoretical findings and demonstrate the efficacy of the proposed framework.
Finally, we conclude the paper in Section~\ref{sec:conclusion}.

In this paper, we use $x \sim \mathcal{CN}(a,b)$ and $x \sim \mathcal{G}(\alpha,\theta)$ to denote a complex Gaussian distribution with mean $a$ and variance $b$, and a gamma distribution with shape $\alpha$ and scale $\theta$, respectively.
We denote the transpose and Hermitian (conjugate) transpose of a vector $\mathbf{x}$ by $\mathbf{x}^T$ and $\mathbf{x}^H$, respectively.
Table~\ref{tab:notation} summarizes the key notations used in this paper.

\section{Network Model} \label{sec:network_model}

We consider a wireless covert network consisting of multiple legitimate nodes and $J$ passive wardens $\mathcal{W} = \{W_1, \ldots, W_J\}$.
The network aims to route sensitive information from a source node~$S$ to a destination node~$D$ with the aid of legitimate relay nodes, while preventing the wardens from detecting the presence of the multi-hop communication.
All legitimate nodes are equipped with $M$ communication modalities.
Each modality $m$ operates at a unique center frequency $f_{m}$ for $m=1,\dots,M$, resulting in distinct channel characteristics.
We define $\mathcal{M} = \{ m_1, \dots, m_M\}$ as the set of all available modalities.
Each warden $W_j$ attempts to detect the presence of communication by measuring the received signals across the frequency ranges of all $M$ modalities.

We consider two warden models based on the degree of coordination among the wardens: 
(i) \textit{Colluding wardens:} All $J$ wardens share their raw observations through a central fusion center, which performs joint hypothesis testing. 
This represents the worst-case warden model. 
(ii) \textit{Non-Colluding wardens:} Each warden $W_j$ independently performs a binary hypothesis test. 
In this case, the overall covertness is determined by the most capable warden.

Let $\Psi$ denote the set of all possible loop-free routes from source~$S$ to destination~$D$. 
For a specific route $\psi \in \Psi$, let $\mathcal{T}_{\psi} = \{ u \mid (u, v) \in \psi \}$ represent the set of transmitters where $(u, v)$ is the single-hop link between transmitter $u$ and receiver $v$.
Each transmitter $u \in \mathcal{T}_{\psi}$ selects a communication modality $m_u \in \mathcal{M}$ for data transmission. 
We assume a block-fading channel model where time is divided into blocks, each consisting of $L$ channel uses. 
The channel coefficients remain constant within each block but vary independently across blocks.

We assume equal \textit{a priori} probabilities for transmission and silence at the source node~$S$, i.e., $\mathbb{P}(\mathcal{H}_0) = \mathbb{P}(\mathcal{H}_1) = 0.5$, where $\mathcal{H}_0$ denotes the null hypothesis (no transmission) and $\mathcal{H}_1$ denotes the alternative hypothesis (transmission is present).
We adopt a simultaneous transmission scheme: under $\mathcal{H}_1$, all transmitters $u \in \mathcal{T}_{\psi}$ along the active route transmit concurrently. 
Consequently, the transmission state is synchronized, resulting in a transmission probability of 0.5 for all active nodes.
To mitigate inter-hop interference arising from this concurrent transmission, we assume the use of code-division multiplexing with orthogonal codes, ensuring that links operating on the same modality remain orthogonal at their respective legitimate receivers. 
Furthermore, we assume that self-interference is effectively suppressed at each transmitter through established self-interference cancellation techniques.

\begin{table}[ht]
\centering
\caption{List of Notations}
\label{tab:notation}
\begin{tabular}{@{}ll@{}}
\toprule
\textbf{Symbol} & \textbf{Definition} \\ \midrule
$J$ & Number of wardens \\
$\mathcal{W}$ & Set of wardens \\
$\Psi$ & Set of all loop-free routes from $S$ to $D$\\
$\psi$ & A specific route $\psi \in \Psi$ \\
$\mathcal{T}_{\psi}$ & Set of active transmitters in route $\psi$ \\
$\mathcal{M}$ & Set of available modalities \\
$M$ & Number of available modalities \\
$m_u$ & Modality selected by node $u$ ($m_u \in \mathcal{M}$) \\
$\mathcal{M}_{\psi}$ & Set of active modalities in $\psi$, $\mathcal{M}_{\psi} = \{ m_u \mid u \in \mathcal{T}_{\psi} \}$ \\
$L$ & Blocklength (channel uses) \\
$\Omega_m$ & Bandwidth of modality $m$ \\
$N_{0,W_j}$ & Noise power spectral density at warden $W_j$ \\
$N_{0,v}$ & Noise power spectral density at receiver $v$ \\
$h_{u,v,m}$ & Channel from node $u$ to node $v$ on modality $m$ \\
$h_{u,W_j,m}$ & Channel from node $u$ to warden $W_j$ on modality $m$ \\
$P_{u}$ & Transmit power of node $u$ \\
$P_{\text{max}}$ & Maximum available power budget at nodes \\
$\rho_{u,v,m_u}$ & SNR at receiver $v$ for link $(u,v)$ on modality $m$ \\
$\rho_{u,W_j,m}$ & SNR at warden $W_j$ from transmitter $u$ on modality $m$ \\
$\rho_{W_j,m}^{(\text{agg})}$ & Aggregate SNR at warden $W_j$ on modality $m$ \\
$\tilde{\mathbf{R}}_m$ & Whitened SNR matrix for modality $m$ \\
$\lambda_{m,j}$ & $j$-th eigenvalue of $\tilde{\mathbf{R}}_m$ \\
$R_{u,v}$ & Achievable rate of link $(u,v)$ \\
$R_{\psi}$ & End-to-end rate of route $\psi$ \\
$R_{\text{req}}$ & Rate requirement \\
$\epsilon$ & Maximum decoding error probability \\
$P_{\text{DEP}}$ & Detection error probability \\ \bottomrule
\end{tabular}
\end{table}

\subsection{Detection Model at Wardens}

We establish a worst-case scenario from a covertness perspective by granting each warden $W_j$ knowledge of the route, channel state information, transmit powers, bandwidths, and modality allocations. 
This assumption ensures that our analysis yields a robust lower bound on covertness performance.

Let $x_u[l] \sim \mathcal{CN}(0,1)$ denote the data symbol transmitted by transmitter $u \in \mathcal{T}_{\psi}$ in channel use $l$, where $\{x_u[l]\}$ are assumed to be independent across transmitters.
The received signal at warden $W_j$ ($j = 1, \ldots, J$) in modality $m$ and channel use $l$ is then given by
\begin{align} 
    &\mathcal{H}_0:  y_{W_j,m}[l] = n_{W_j,m}[l], \quad \forall m \in \mathcal{M}, \label{eq:Received_signal_Willie} \\ \nonumber
    &\mathcal{H}_1:  y_{W_j,m}[l] = 
    \begin{cases}
        \displaystyle &\!\!\! \sum_{u \in \mathcal{T}_{\psi}} \mathbb{I}(m_u = m) \sqrt{P_{u}} h_{u,W_j,m} x_u[l] \\ 
        &\quad + n_{W_j,m}[l],  ~~\text{if}~~ m \in \mathcal{M}_{\psi}, \\
        &\!\!\! n_{W_j,m}[l], ~~\text{otherwise},
    \end{cases} 
\end{align} 
where  $\mathbb{I}(\cdot)$ denotes the indicator function, which equals 1 if the condition inside the brackets is satisfied and 0 otherwise.
Here, $\mathcal{M}_{\psi} = \{ m_u \mid u \in \mathcal{T}_{\psi} \}$ represents the subset of modalities active in the current route $\psi$.
Under $\mathcal{H}_0$, the network is silent, and warden~$W_j$ observes only the AWGN component $n_{W_j,m}[l] \sim \mathcal{CN}(0, N_{0,W_j} \Omega_m)$, where $N_{0,W_j}$ and $\Omega_m$ denote the noise power spectral density at warden $W_j$ and the allocated bandwidth of modality $m$, respectively.
Under $\mathcal{H}_1$, for any active modality $m \in \mathcal{M}_{\psi}$, warden~$W_j$ observes the superposition of signals from all nodes transmitting on modality~$m$, where $P_{u}$ is the transmit power of transmitter~$u$ and $h_{u,W_j,m}$ is the channel coefficient from transmitter~$u$ to warden~$W_j$ on modality~$m$.

The signal-to-noise ratio (SNR) at warden~$W_j$ for the individual link $(u,v)$ on modality~$m_u$ is defined as
\begin{equation} \label{eq:indv_SNR_W_def}
    \rho_{u,W_j,m_u} = \frac{P_{u} |h_{u,W_j,m_u}|^2}{\Omega_{m_u} N_{0,W_j}}.
\end{equation}
Accordingly, the aggregate SNR at warden~$W_j$ on modality~$m$ is obtained by summing the individual SNRs from all transmitters using that modality:
\begin{equation} \label{eq:agg_SNR_W_def}
    \rho_{W_j,m}^{(\text{agg})} = \sum_{u \in \mathcal{T}_{\psi}} \mathbb{I}(m_u = m) \rho_{u, W_j, m}. 
\end{equation}

For the \textit{colluding} case, all $J$ wardens forward their raw observations to a central fusion center that renders a joint decision based on the collected signals.
The corresponding detection error probability (DEP) is defined as~\cite{Chen:23}
\begin{equation} \label{eq:DEP_coll_def}
    P_{\text{DEP},\psi}^{(\text{coll})} = P_{\text{FA},\psi}^{(\text{coll})} + P_{\text{MD},\psi}^{(\text{coll})},
\end{equation}
where $P_{\text{FA},\psi}^{(\text{coll})} = \mathbb{P}(\mathcal{D}_1 | \mathcal{H}_0)$ and $P_{\text{MD},\psi}^{(\text{coll})} = \mathbb{P}(\mathcal{D}_0 | \mathcal{H}_1)$ are the false alarm and missed detection probabilities, respectively, with $\mathcal{D}_0$ and $\mathcal{D}_1$ denoting the decisions in favor of $\mathcal{H}_0$ and $\mathcal{H}_1$.

For the \textit{non-colluding} case, the DEP is governed by the most capable warden.
Specifically, the DEP is given by
\begin{equation} \label{eq:DEP_nc_def}
    P_{\text{DEP},\psi}^{(\text{nc})} = \min_{j=1,\dots,J} (P_{\text{FA},\psi,j} + P_{\text{MD},\psi,j}),
\end{equation}
where $P_{\text{FA},\psi,j} = \mathbb{P}(\mathcal{D}_{1,j} | \mathcal{H}_0)$ and $P_{\text{MD},\psi,j} = \mathbb{P}(\mathcal{D}_{0,j} | \mathcal{H}_1)$ represent the individual false alarm and missed detection probabilities at warden~$W_j$, respectively, and $\mathcal{D}_{i,j}$ denotes the decision of warden~$W_j$ in favor of $\mathcal{H}_i$.

\subsection{Communication Performance}

For a specific link $(u, v)$ belonging to the active route~$\psi$, the received signal at node~$v$ via modality~$m_u$ selected by transmitter~$u$ is given by
\begin{equation} \label{eq:received_signal_rx}
y_{v, m_u}[l] = \sqrt{P_u} h_{u, v, m_u} x_u[l] + n_{v, m_u}[l], \quad l = 1, \dots, L,
\end{equation}
where $h_{u, v, m_u}$ is the channel coefficient from~$u$ to~$v$ on modality~$m_u$ and $n_{v, m_u}[l] \sim \mathcal{CN}(0, N_{0,v} \Omega_{m_u})$ represents the AWGN at receiver~$v$ with noise power spectral density $N_{0,v}$.
The SNR at node~$v$ is $\rho_{u,v,m_u} = \frac{P_{u} |h_{u,v,m_u}|^2}{N_{0,v} \Omega_{m_u}}$.
The achievable rate $R_{u,v}$ (in bits/sec) for link $(u,v) \in \psi$ subject to a maximum decoding error probability~$\epsilon$ is given by~\cite{Yan:19b}
\begin{align} \label{eq:rate_hop}
&R_{u,v}(m_u, P_u) \\
&=  \Omega_{m_u}  \Bigg[  \! \log_2\left(1 \!+\! \rho_{u,v,m_u}\right) - \frac{K}{\ln(2)} \!\sqrt{1 - (1+\rho_{u,v,m_u})^{-2}  }   \Bigg], \nonumber 
\end{align}
where $K \triangleq \frac{Q^{-1}(\epsilon)}{\sqrt{L}}$ and $Q^{-1}(\cdot)$ denotes the inverse Q-function. 
The end-to-end rate of route~$\psi$ is therefore expressed as
\begin{equation} \label{eq:rate_route}
R_{\psi} = \min_{(u,v) \in \psi} R_{u,v}(m_u, P_u).
\end{equation}

\subsection{Optimization Problem}

We investigate the joint optimization of routing, modality selection, and transmit power to maximize the DEP at the wardens while satisfying a minimum required data rate at the destination:
\begin{align} \label{eq:opt_problem}
\max_{\psi, \{m_u\}, \{P_u\}} & \quad P_{\text{DEP},\psi} \left( \{m_u\}, \{P_u\} \right) \\  \label{eq:constraint_power}
            \text{s.t.} & \quad P_u \leq P_{\text{max}}, \forall u \in \mathcal{T}_{\psi}, \\ \label{eq:constraint_rate}
                        & \quad R_{\psi} \geq R_{\text{req}},  
\end{align}
where $P_{\text{max}}$ denotes the maximum available power at transmitters and the DEP $P_{\text{DEP},\psi}$ is defined for each warden model in Sections~\ref{sec:colluding_analysis} and~\ref{sec:noncolluding_analysis}.

\section{DEP Analysis: Colluding Wardens} \label{sec:colluding_analysis}

In this section, we derive the optimal detection strategy and analyze the DEP for the case where all $J$ wardens pool their raw observations at a central fusion center that performs joint hypothesis testing.
In addition, we present a low-complexity closed-form approximation of the DEP, a tractable lower bound on the DEP, and an alternative DEP formulation under per-hop detection.

\subsection{Optimal Decision Rule and Exact DEP Analysis}

When all $J$ wardens collude, the fusion center observes a $J$-dimensional received signal vector $\mathbf{y}_m[l] = \big[ y_{W_1,m}[l],  \dots,  y_{W_J,m}[l] \big]^T \in \mathbb{C}^{J}$ for each modality $m \in \mathcal{M}_{\psi}$ and channel use~$l$.
Under hypotheses $\mathcal{H}_0$ and $\mathcal{H}_1$, respectively, the distribution of $\mathbf{y}_m[l]$ is 
\begin{align}
    &\mathcal{H}_0: \quad \mathbf{y}_m[l] \sim \mathcal{CN}(\mathbf{0}, \mathbf{C}_{0,m}), \label{eq:vector_H0} \\
    &\mathcal{H}_1: \quad \mathbf{y}_m[l] \sim \mathcal{CN}(\mathbf{0}, \mathbf{C}_{1,m}), \label{eq:vector_H1}
\end{align}
where the noise covariance under $\mathcal{H}_0$ is
\begin{equation} \label{eq:C0}
    \mathbf{C}_{0,m} = \text{diag}\left(\Omega_m N_{0,W_1}, \ldots, \Omega_m N_{0,W_J}\right),
\end{equation}
and the signal-plus-noise covariance under $\mathcal{H}_1$ is
\begin{equation} \label{eq:C1}
    \mathbf{C}_{1,m} = \mathbf{C}_{0,m} + \sum_{u \in \mathcal{T}_{\psi}} \mathbb{I}(m_u = m) P_u \, \mathbf{h}_{u,m}\mathbf{h}_{u,m}^H,
\end{equation}
where $\mathbf{h}_{u,m} = [h_{u,W_1,m}, \ldots, h_{u,W_J,m}]^T \in \mathbb{C}^{J}$ is the channel vector from transmitter $u$ to all wardens.

To diagonalize the detection problem, we apply the whitening transformation $\mathbf{z}_m[l] = \mathbf{C}_{0,m}^{-1/2} \mathbf{y}_m[l]$, yielding
\begin{align}
    &\mathcal{H}_0: \quad \mathbf{z}_m[l] \sim \mathcal{CN}(\mathbf{0}, \mathbf{I}_J), \\
    &\mathcal{H}_1: \quad \mathbf{z}_m[l] \sim \mathcal{CN}(\mathbf{0}, \mathbf{I}_J + \tilde{\mathbf{R}}_m),
\end{align}
where $\mathbf{I}_J$ denotes the $J \times J$ identity matrix and the whitened SNR matrix $\tilde{\mathbf{R}}_m$ is defined as
\begin{align} \label{eq:whitened_SNR}
    \tilde{\mathbf{R}}_m &\triangleq \mathbf{C}_{0,m}^{-1/2} \left( \sum_{u \in \mathcal{T}_{\psi}} \mathbb{I}(m_u = m) P_u \, \mathbf{h}_{u,m}\mathbf{h}_{u,m}^H \right) \mathbf{C}_{0,m}^{-1/2} \nonumber \\
    &= \sum_{u \in \mathcal{T}_{\psi}} \mathbb{I}(m_u = m) P_u \, \tilde{\mathbf{h}}_{u,m}\tilde{\mathbf{h}}_{u,m}^H,
\end{align}
with the normalized channel vector $\tilde{\mathbf{h}}_{u,m} = \mathbf{C}_{0,m}^{-1/2} \mathbf{h}_{u,m}$, whose $j$-th entry is $\frac{h_{u,W_j,m}}{\sqrt{\Omega_m N_{0,W_j}}}$.

Let $\{\lambda_{m,j}\}_{j=1}^{J}$ denote the eigenvalues of $\tilde{\mathbf{R}}_m$ and let $\mathbf{U}_m$ be the corresponding unitary eigenvector matrix such that $\tilde{\mathbf{R}}_m = \mathbf{U}_m \boldsymbol{\Lambda}_m \mathbf{U}_m^H$ where $\boldsymbol{\Lambda}_m = \text{diag}(\lambda_{m,1}, \ldots, \lambda_{m,J})$.
By projecting onto the eigenbasis via $\mathbf{q}_m[l] = \mathbf{U}_m^H \mathbf{z}_m[l] = [q_{m,1}[l], \ldots, q_{m,J}[l]]^T$, we obtain the independent components $q_{m,j}[l]$ distributed as 
\begin{align}
    &\mathcal{H}_0: \quad q_{m,j}[l] \sim \mathcal{CN}(0, 1), \label{eq:eigen_H0} \\
    &\mathcal{H}_1: \quad q_{m,j}[l] \sim \mathcal{CN}(0, 1 + \lambda_{m,j}). \label{eq:eigen_H1}
\end{align}

Consequently, by defining $\mathbf{q}$ as the collection of all observations $q_{m,j}[l]$ over all channel uses, active modalities, and wardens, the likelihood ratio test (LRT) is written as
\begin{align} \label{eq:LRT_colluding}
	\Lambda^{(\text{coll})} = \frac{f(\mathbf{q} |\mathcal{H}_1)}{f(\mathbf{q} |\mathcal{H}_0)} \underset{\mathcal{D}_0}{\overset{\mathcal{D}_1}{\gtrless}}  1.
\end{align}
The corresponding log-likelihood ratio (LLR) is derived as follows:
\begin{align} \label{eq:LLR}
    &\ln\left( \Lambda^{(\text{coll})} \right) = \ln \left( \frac{\prod_{m \in \mathcal{M}_{\psi}} \prod_{j=1}^{J} \prod_{l=1}^{L} p(q_{m,j}[l] | \mathcal{H}_1)}{\prod_{m \in \mathcal{M}_{\psi}} \prod_{j=1}^{J} \prod_{l=1}^{L} p(q_{m,j}[l] | \mathcal{H}_0)} \right) \nonumber \\
    &= \sum_{m, j, l} \ln \left( \frac{\frac{1}{\pi(1+\lambda_{m,j})} \exp\left( -\frac{|q_{m,j}[l]|^2}{1+\lambda_{m,j}} \right)}{\frac{1}{\pi} \exp\left( -|q_{m,j}[l]|^2 \right)} \right) \nonumber \\
    &= \sum_{m, j, l} \left[ \ln\left(\frac{1}{1+\lambda_{m,j}}\right) - \frac{|q_{m,j}[l]|^2}{1+\lambda_{m,j}} + |q_{m,j}[l]|^2 \right] \nonumber \\
    &= \sum_{m, j, l} \left[ |q_{m,j}[l]|^2 \left( 1 - \frac{1}{1+\lambda_{m,j}} \right) - \ln(1+\lambda_{m,j}) \right] \nonumber \\
    &= \sum_{m, j} \underbrace{\frac{\lambda_{m,j}}{1+\lambda_{m,j}}}_{w_{m,j}} \underbrace{\sum_{l=1}^{L} |q_{m,j}[l]|^2}_{E_{m,j}} - \underbrace{\sum_{m, j} L \ln(1+\lambda_{m,j})}_{\delta_{\psi}^{(\text{coll})}}.
\end{align}
Finally, from~\eqref{eq:LRT_colluding} and~\eqref{eq:LLR},  the optimal decision rule that minimizes the DEP is given by
\begin{equation} \label{eq:test_statistic_coll}
    T_{\psi}^{(\text{coll})} \triangleq \sum_{m \in \mathcal{M}_{\psi}} \sum_{j=1}^{J} w_{m,j} E_{m,j} \underset{\mathcal{D}_0}{\overset{\mathcal{D}_1}{\gtrless}} \delta_{\psi}^{(\text{coll})}.
\end{equation}

\begin{remark} \label{remark:eigenvalue_single_warden}
The colluding detector generalizes the single-warden $W_j$ detector by replacing $\lambda_{m,j}$ with $\rho_{W_j,m}^{(\text{agg})}$ for all $m$, since the matrix $\tilde{\mathbf{R}}_m$ reduces to a scalar $\rho_{W_j,m}^{(\text{agg})}$ when $J=1$.
\end{remark}

To evaluate the exact covertness performance, we derive the distribution of the test statistic $T_{\psi}^{(\text{coll})}$ given in~\eqref{eq:test_statistic_coll}.
Under hypothesis $\mathcal{H}_i$ ($i \in \{0,1\}$), the independent components follow $q_{m,j}[l] \sim \mathcal{CN}(0, \sigma_{i,m,j}^2)$, where the variances are given by $\sigma_{0,m,j}^2 = 1$ and $\sigma_{1,m,j}^2 = 1 + \lambda_{m,j}$.
The energy term $E_{m,j} = \sum_{l=1}^{L} |q_{m,j}[l]|^2$ represents a sum of $L$ independent squared magnitudes of complex Gaussian variables. Consequently, $E_{m,j}$ follows a gamma distribution with shape parameter $L$ and scale parameter $\sigma_{i,m,j}^2$, denoted by $E_{m,j} \sim \mathcal{G}(L, \sigma_{i,m,j}^2)$.
By the scaling property of the gamma distribution, the weighted term in the test statistic follows $w_{m,j} E_{m,j} \sim \mathcal{G}(L, w_{m,j} \sigma_{i,m,j}^2)$.

Since the test statistic $T_{\psi}^{(\text{coll})}$ is a sum of independent random variables across all active modalities and wardens, we utilize the characteristic function (CF) for its distributional analysis. The CF of a gamma-distributed variable $X \sim \mathcal{G}(k, \theta)$ is $\phi_X(t) = (1 - \mathrm{j} t \theta)^{-k}$, where $\mathrm{j} = \sqrt{-1}$. Therefore, the CF of $T_{\psi}^{(\text{coll})}$ under hypothesis $\mathcal{H}_i$ is the product of the individual CFs:
\begin{equation} \label{eq:CF_coll}
    \phi_{T^{(\text{coll})} | \mathcal{H}_i}(t) = \prod_{m \in \mathcal{M}_{\psi}} \prod_{j=1}^{J} \left(1 - \mathrm{j} t \, w_{m,j} \sigma_{i,m,j}^2 \right)^{-L}.
\end{equation}

To compute the error probabilities from the CFs, we employ the Gil-Pelaez inversion formula~\cite{GIL-PELAEZ:51}, which expresses the cumulative distribution function (CDF) $F(x)$ of a random variable $X$ with CF $\phi_X(t)$ as $F(x) = \mathbb{P}(X \le x) = \frac{1}{2} - \frac{1}{\pi} \int_0^\infty \frac{1}{t} \Im \left\{ e^{-\mathrm{j}tx} \phi_X(t) \right\} dt$.
From $P_{\text{DEP},\psi}^{(\text{coll})}$ in~\eqref{eq:DEP_coll_def} and the optimal decision rule in~\eqref{eq:test_statistic_coll}, we express the exact DEP as
\begin{align} \label{eq:DEP_exact_coll}
    &P_{\text{DEP},\psi}^{(\text{coll})} \!=\! \mathbb{P} \! \left(T_{\psi}^{(\text{coll})} \!>\! \delta_{\psi}^{(\text{coll})} | \mathcal{H}_0 \! \right) \!+\! \mathbb{P}\!\left(T_{\psi}^{(\text{coll})} \!\le\! \delta_{\psi}^{(\text{coll})} | \mathcal{H}_1 \! \right) \\
    &= \left( 1 - F_{T_{\psi}^{(\text{coll})}|\mathcal{H}_0}\left(\delta_{\psi}^{(\text{coll})}\right) \right) + F_{T_{\psi}^{(\text{coll})}|\mathcal{H}_1}\left(\delta_{\psi}^{(\text{coll})}\right) \nonumber \\
    &= 1 \!+\! \frac{1}{\pi} \int_{0}^{\infty} \!\! \frac{1}{t} \Im \left\{ e^{-\mathrm{j} t \delta_{\psi}^{(\text{coll})}} \!\! \left( \phi_{T^{(\text{coll})} | \mathcal{H}_0}(t) \!-\! \phi_{T^{(\text{coll})} | \mathcal{H}_1}(t) \right) \right\} dt.\nonumber 
\end{align}

\subsection{Closed-Form DEP Approximation Via Moment Matching} 
\label{subsec:approx_dep_coll}

While the exact DEP calculation in~\eqref{eq:DEP_exact_coll} involves numerical integration, a practical closed-form approximation can be obtained by matching the first two moments (mean and variance) of the test statistic to a single gamma distribution.
Recall that the test statistic is a weighted sum of independent gamma-distributed variables $T_{\psi}^{(\text{coll})} = \sum_{m \in \mathcal{M}_{\psi}} \sum_{j=1}^{J} w_{m,j} E_{m,j}$ where $E_{m,j} \sim \mathcal{G}(L, \sigma_{i,m,j}^2)$ under hypothesis $\mathcal{H}_i$.

First, we compute the moments of the individual terms.
Using the properties of the gamma distribution, the mean and variance of the weighted term $w_{m,j} E_{m,j}$ are
\begin{align}
    \mathbb{E}[w_{m,j} E_{m,j} | \mathcal{H}_i] &=  L w_{m,j}  \sigma_{i,m,j}^2, \\
    \text{Var}(w_{m,j} E_{m,j} | \mathcal{H}_i) &= L w_{m,j}^2 (\sigma_{i,m,j}^2)^2.
\end{align}
Since the components are independent across modalities and eigenmodes, the total mean $\mu_{i}^{(\text{coll})}$ and variance $V_{i}^{(\text{coll})}$ of $T_{\psi}^{(\text{coll})}$ are the sums of the individual means and variances:
\begin{align}
    \mu_{i}^{(\text{coll})} &= \sum_{m \in \mathcal{M}_{\psi}} \sum_{j=1}^{J} L \, w_{m,j} \sigma_{i,m,j}^2, \label{eq:mean_coll} \\
    V_{i}^{(\text{coll})} &= \sum_{m \in \mathcal{M}_{\psi}} \sum_{j=1}^{J} L \left( w_{m,j} \sigma_{i,m,j}^2 \right)^2. \label{eq:var_coll}
\end{align}

We approximate the distribution of $T_{\psi}^{(\text{coll})}$ under $\mathcal{H}_i$ by a single gamma distribution $\tilde{T}^{(\text{coll})} \sim \mathcal{G}(\tilde{\alpha}_{i}^{(\text{coll})}, \tilde{\theta}_{i}^{(\text{coll})})$.
By equating the moments of the approximation to the exact moments, the shape and scale parameters are obtained as
\begin{equation} \label{eq:approx_params_coll}
    \tilde{\alpha}_{i}^{(\text{coll})} = \frac{\left(\mu_{i}^{(\text{coll})}\right)^2}{V_{i}^{(\text{coll})}}, \quad \tilde{\theta}_{i}^{(\text{coll})} = \frac{V_{i}^{(\text{coll})}}{\mu_{i}^{(\text{coll})}}.
\end{equation}

The error probabilities are then approximated using the CDF of the gamma distribution, $F(x; \alpha, \theta) = \frac{\gamma(\alpha, x/\theta)}{\Gamma(\alpha)}$, where $\Gamma(s)=\int_0^\infty t^{s-1} e^{-t} dt$ and $\gamma(s,x) = \int_0^x t^{s-1} e^{-t} dt$ denote the gamma function and the lower incomplete gamma function, respectively.
The false alarm and missed detection probabilities are approximated as $P_{\text{FA}}^{(\text{coll})} \approx 1 - F(\delta_{\psi}^{(\text{coll})}; \tilde{\alpha}_0, \tilde{\theta}_0)$ and  $P_{\text{MD}}^{(\text{coll})} \approx F(\delta_{\psi}^{(\text{coll})}; \tilde{\alpha}_1, \tilde{\theta}_1)$, respectively.
Summing these yields the approximate DEP in closed form as
\begin{equation} \label{eq:DEP_approx_coll}
    P_{\text{DEP},\psi}^{(\text{coll})} \! \approx \! 1 \!-\! \frac{\gamma\!\left( \!\tilde{\alpha}_{0}^{(\text{coll})},\, \delta_{\psi}^{(\text{coll})} \!/ \tilde{\theta}_{0}^{(\text{coll})} \!\right)}{\Gamma\!\left(\tilde{\alpha}_{0}^{(\text{coll})}\right)} + \frac{\gamma\!\left( \! \tilde{\alpha}_{1}^{(\text{coll})},\, \delta_{\psi}^{(\text{coll})} \!/ \tilde{\theta}_{1}^{(\text{coll})} \! \right)}{\Gamma\!\left(\tilde{\alpha}_{1}^{(\text{coll})}\right)}.
\end{equation}

\subsection{KL Divergence-Based DEP Lower Bound} \label{subsec:KL_coll}

From Pinsker's inequality, the DEP is lower bounded by the relative entropy (KL divergence $\mathcal{D}^{(\text{coll})}(\mathbb{P}_0 \| \mathbb{P}_1)$) between the distributions under the two hypotheses as 
\begin{equation} \label{eq:DEP_LB_coll_def}
    P_{\text{DEP},\psi}^{(\text{coll})} \geq 1 - \sqrt{\frac{1}{2} \mathcal{D}^{(\text{coll})}(\mathbb{P}_0 \| \mathbb{P}_1)}.
\end{equation}
Recall from~\eqref{eq:eigen_H0} and~\eqref{eq:eigen_H1} that the detection problem has been diagonalized into independent components $q_{m,j}[l]$.
Due to the independence of observations across time slots $l$, modalities $m$, and eigenmodes $j$, the total KL divergence decomposes as the sum of the KL divergences of the individual scalar components: 
\begin{align}
    &\mathcal{D}^{(\text{coll})}(\mathbb{P}_0 \| \mathbb{P}_1) \nonumber \\ 
    &=  \sum_{m \in \mathcal{M}_{\psi}} \sum_{j=1}^{J} \sum_{l=1}^{L} \mathcal{D}(p(q_{m,j}[l]|\mathcal{H}_0) \| p(q_{m,j}[l]|\mathcal{H}_1)),  
\end{align}
where $p(q_{m,j}[l]|\mathcal{H}_i)$ is the probability density function (PDF) of $q_{m,j}[l]$ under hypothesis $\mathcal{H}_i$.

Since $q_{m,j}[l] \sim \mathcal{CN}(0, 1)$ under $\mathcal{H}_0$ and  $q_{m,j}[l] \sim \mathcal{CN}(0, 1 + \lambda_{m,j})$ under $\mathcal{H}_1$, the per-component KL divergence evaluates to
\begin{equation}
    \mathcal{D}(p(q_{m,j}[l]|\mathcal{H}_0) \| p(q_{m,j}[l]|\mathcal{H}_1)) = \ln(1 \!+\! \lambda_{m,j}) \!-\! \frac{\lambda_{m,j}}{1 \!+\! \lambda_{m,j}}.
\end{equation}
Consequently, the total KL divergence becomes
\begin{equation} \label{eq:KL_coll}
    \mathcal{D}^{(\text{coll})}(\mathbb{P}_0 \| \mathbb{P}_1) \!=\! L \!\! \sum_{m \in \mathcal{M}_{\psi}} \! \sum_{j=1}^{J}  \bigg[  \ln(  1 \!+\! \lambda_{m,j}) \!-\! \frac{\lambda_{m,j}}{1 \!+\! \lambda_{m,j}} \bigg],
\end{equation}
and the corresponding lower bound is given by
\begin{equation} \label{eq:DEP_LB_coll}
    P_{\text{DEP},\psi}^{(\text{coll})} \!\geq\! 1 \!-\! \sqrt{\frac{L}{2} \!\!  \sum_{m \in \mathcal{M}_{\psi}} \sum_{j=1}^{J} \left[ \ln(1 \!+\! \lambda_{m,j}) \!-\! \frac{\lambda_{m,j}}{1 \!+\! \lambda_{m,j}} \right]        }.
\end{equation}

\begin{remark}
It is important to distinguish the detection model considered in this work from the prior multi-hop covert routing study in~\cite{Sheikholeslami:18}, which assumes that wardens observe transmissions from different hops independently.
In contrast, our work considers a \textit{simultaneous transmission} scheme in which the wardens observe the coherent superposition of signals from all active transmitters. 
This represents a worst-case adversarial scenario in which the colluding wardens can exploit receive beamforming gains  and spatial correlations arising from simultaneous transmission, thereby achieving a lower DEP than what independent per-hop link metric summation would predict.
\end{remark}

\subsection{DEP formulation under Per-Hop Detection} \label{subsec:prod_DEP_coll}

Following the framework of~\cite{Kong:24TIFS}, we consider a detection model where the wardens attempt to detect transmissions on a per-hop basis.
Specifically, the wardens are assumed to perform detection by processing the signal from each active link $(u, v) \in \psi$ in isolation\footnote{Note that the per-hop detection model is evaluated as a theoretical baseline to facilitate comparison with single-hop routing metrics, rather than as a practical adversarial strategy.}. 
Under this model, each hop $(u,v)$ is detected by all $J$ colluding wardens.

For any given hop $(u,v)$, the colluding wardens employ maximal ratio combining to maximize detection performance. 
Consequently, the effective SNR becomes the sum of the SNRs at all individual wardens:
\begin{equation} \label{eq:rho_coll_hop}
    \rho_{u}^{(\text{coll})} \triangleq \sum_{j=1}^{J} \rho_{u,W_j,m_u} = \sum_{j=1}^{J} \frac{P_u |h_{u,W_j,m_u}|^2}{\Omega_{m_u} N_{0,W_j}}.
\end{equation}
The DEP of each hop $(u,v)$ is then given by~\cite{Kong:24TIFS}:
\begin{align} \label{eq:DEP_per_hop_coll}
    P_{\text{DEP},u,v}^{(\text{coll})} = 1 - & \frac{1}{\Gamma(L)} \Bigg[ \gamma\!\left(L,\, L\!\left(1 + \frac{1}{\rho_u^{(\text{coll})}}\right) \ln\!\left(1 + \rho_u^{(\text{coll})}\right)\right) \nonumber  \\
    &- \gamma\!\left(L,\, \frac{L}{\rho_u^{(\text{coll})}} \ln\!\left(1 + \rho_u^{(\text{coll})}\right)\right) \Bigg].
\end{align}
Finally, the end-to-end DEP is modeled as the product of the individual link DEPs:
\begin{equation} \label{eq:DEP_product_coll}
    P_{\text{DEP},\psi}^{(\text{coll,prod})} = \prod_{(u,v) \in \psi} P_{\text{DEP},u,v}^{(\text{coll})}.
\end{equation}

We note that the product-based DEP formulation in~\eqref{eq:DEP_product_coll} yields a lower value than the exact colluding DEP derived in~\eqref{eq:DEP_exact_coll}. 
This is because the product of per-hop DEPs diminishes geometrically with path length, as each additional link contributes a multiplicative factor strictly less than unity, causing the product to decrease monotonically with the number of hops.

\section{DEP Analysis: Non-Colluding Wardens} \label{sec:noncolluding_analysis}

In this section, we analyze the DEP for the case where each warden $W_j$ independently performs a binary hypothesis test on its own observations.
The overall network-level DEP is then governed by the worst-case (most capable) individual warden.

\subsection{Exact DEP Analysis}

Since each warden $W_j$ operates independently, its detection structure is identical to the single-warden case, i.e., $J = 1$.
Under hypothesis $\mathcal{H}_i$ ($i \in \{0,1\}$), from~\eqref{eq:Received_signal_Willie}, the received signal at warden $W_j$ in modality $m \in \mathcal{M}_{\psi}$ follows
\begin{equation}
    \mathcal{H}_i: \quad y_{W_j,m}[l] \sim \mathcal{CN}\left( 0, \sigma_{i,m,j}^2 \right),
\end{equation}
where $\sigma_{0,m,j}^2 = \Omega_m N_{0,W_j}$ and $\sigma_{1,m,j}^2 = \Omega_m N_{0,W_j}\big(1 + \rho_{W_j,m}^{(\text{agg})} \big)$, with the aggregated SNR at warden $W_j$ on modality $m$ as defined in~\eqref{eq:agg_SNR_W_def}.
Then, from~\eqref{eq:LLR}, \eqref{eq:test_statistic_coll}, and Remark~\ref{remark:eigenvalue_single_warden}, the optimal detector at warden $W_j$ that minimizes its individual DEP is
\begin{equation} \label{eq:test_statistic_nc}
    T_{\psi,j} \triangleq \sum_{m \in \mathcal{M}_{\psi}} w_{m,j} E_{m,j}  \underset{\mathcal{D}_0}{\overset{\mathcal{D}_1}{\gtrless}} \delta_{\psi,j},
\end{equation}
where $E_{m,j} = \sum_{l=1}^{L} |y_{W_j,m}[l]|^2$, $w_{m,j} = \frac{\rho_{W_j,m}^{(\text{agg})}}{\sigma_{0,m,j}^2 \left(1 + \rho_{W_j,m}^{(\text{agg})} \right)}$, and $\delta_{\psi,j} = \sum_{m \in \mathcal{M}_{\psi}} L \ln \left(1 + \rho_{W_j,m}^{(\text{agg})} \right)$.

The false alarm probability $P_{\text{FA},\psi,j} = \mathbb{P}( T_{\psi,j}  > \delta_{\psi,j} | \mathcal{H}_0 )$ and missed detection probability $P_{\text{MD},\psi,j} = \mathbb{P}( T_{\psi,j}  \le \delta_{\psi,j} | \mathcal{H}_1 )$ at warden $W_j$ are obtained via the Gil-Pelaez inversion formula~\cite{GIL-PELAEZ:51} as
\begin{align}
    P_{\text{FA},\psi,j} &=  \frac{1}{2} + \frac{1}{\pi} \int_0^{\infty} \frac{1}{t} \Im\!\left\{ e^{-\mathrm{j}t\delta_{\psi,j}} \phi_{T_j|\mathcal{H}_0}(t) \right\} dt, \label{eq:PFA_j_exact} \\
    P_{\text{MD},\psi,j} &= \frac{1}{2} - \frac{1}{\pi} \int_0^{\infty} \frac{1}{t} \Im\!\left\{ e^{-\mathrm{j}t\delta_{\psi,j}} \phi_{T_j|\mathcal{H}_1}(t) \right\} dt, \label{eq:PMD_j_exact}
\end{align}
where the characteristic function of warden $W_j$'s test statistic is
\begin{equation} \label{eq:CF_nc}
    \phi_{T_j|\mathcal{H}_i}(t) = \prod_{m \in \mathcal{M}_{\psi}} \left(1 - \mathrm{j}t \, w_{m,j} \sigma_{i,m,j}^2 \right)^{-L}.
\end{equation}
Consequently, the DEP at warden $W_j$ is given by 
\begin{equation} \label{eq:DEP_per_warden}
    P_{\text{DEP},\psi,j} = P_{\text{FA},\psi,j} + P_{\text{MD},\psi,j}.
\end{equation}
Since the network covertness is determined by the warden with the lowest (worst-case) DEP, the overall DEP for non-colluding wardens is given by 
\begin{equation} \label{eq:DEP_nc}
    P_{\text{DEP},\psi}^{(\text{nc})} = \min_{j=1,\ldots,J} P_{\text{DEP},\psi,j}.
\end{equation}

\subsection{Closed-Form DEP Approximation Via Moment Matching} 
\label{subsec:approx_dep_nc}

For each warden $W_j$, analogously to Section~\ref{subsec:approx_dep_coll}, we approximate $T_{\psi,j}$ under $\mathcal{H}_i$ by a single gamma distribution $\tilde{T}_{\psi,j} \sim \mathcal{G}(\tilde{\alpha}_{i,\psi,j}, \tilde{\theta}_{i,\psi,j})$ with matched moments. The mean and variance of $T_{\psi,j}$ under $\mathcal{H}_i$ are
\begin{align}
    \mu_{i,\psi,j} &= \sum_{m \in \mathcal{M}_{\psi}} L \, w_{m,j} \sigma_{i,m,j}^2, \label{eq:mean_nc} \\
    V_{i,\psi,j} &= \sum_{m \in \mathcal{M}_{\psi}} L \left( w_{m,j} \sigma_{i,m,j}^2 \right)^2. \label{eq:var_nc}
\end{align}
Hence, the matched gamma parameters are $\tilde{\alpha}_{i,\psi,j} = \mu_{i,\psi,j}^2 / V_{i,\psi,j}$ and $\tilde{\theta}_{i,\psi,j} = V_{i,\psi,j} / \mu_{i,\psi,j}$.
The approximate false alarm and missed detection probabilities at warden $W_j$ are then given by
\begin{align}
    &P_{\text{FA},\psi,j} \!\approx\! \mathbb{P}\left( \tilde{T}_{\psi,j}  > \delta_{\psi,j} | \mathcal{H}_0 \right) \!=\!  1 \!-\! \frac{\gamma(\tilde{\alpha}_{0,j},\, \delta_{\psi,j}/\tilde{\theta}_{0,j})}{\Gamma(\tilde{\alpha}_{0,j})}, \label{eq:PFA_approx} \\
    &P_{\text{MD},\psi,j} \!\approx\! \mathbb{P}\left( \tilde{T}_{\psi,j}  \le \delta_{\psi,j} | \mathcal{H}_1 \right) = \frac{\gamma(\tilde{\alpha}_{1,j},\, \delta_{\psi,j}/\tilde{\theta}_{1,j})}{\Gamma(\tilde{\alpha}_{1,j})}. \label{eq:PMD_approx}
\end{align}
By combining \eqref{eq:PFA_approx} and~\eqref{eq:PMD_approx} with~\eqref{eq:DEP_per_warden} and~\eqref{eq:DEP_nc}, we obtain a closed-form approximation of the overall DEP for the non-colluding warden case.

\subsection{KL Divergence-Based DEP Lower Bound} \label{subsec:KL_nc}

For each warden $W_j$, from~\eqref{eq:KL_coll} and Remark~\ref{remark:eigenvalue_single_warden}, the per-warden KL divergence is given by
\begin{equation} \label{eq:KL_nc_j}
    \mathcal{D}_{\psi,j} = L \sum_{m \in \mathcal{M}_{\psi}}  \left[ \ln\left(1 + \rho_{W_j,m}^{(\text{agg})}\right) - \frac{\rho_{W_j,m}^{(\text{agg})}}{1 + \rho_{W_j,m}^{(\text{agg})}} \right].
\end{equation}
By Pinsker's inequality, the per-warden DEP is bounded as $P_{\text{DEP},\psi,j} \geq 1 - \sqrt{\frac{1}{2}\mathcal{D}_{\psi,j}}$.
Consequently, we obtain a lower bound on the overall DEP as
\begin{align} \label{eq:DEP_LB_nc}
    &P_{\text{DEP},\psi}^{(\text{nc})} = \min_{j=1,\ldots,J} P_{\text{DEP},\psi,j} \nonumber \\
    &\geq  \! 1 \!-\! \max_{j=1,\ldots,J}  \sqrt{ \! \frac{L}{2} \!\!  \sum_{m \in \mathcal{M}_{\psi}}  \! \Bigg[ \! \ln \! \left( \! 1 \!+\! \rho_{W_j,m}^{(\text{agg})} \! \right) \!-\! \frac{\rho_{W_j,m}^{(\text{agg})}}{1 \!+\! \rho_{W_j,m}^{(\text{agg})}} \Bigg]        }.
\end{align}

\subsection{DEP formulation under Per-Hop Detection} \label{subsec:prod_DEP_nc}

In the per-hop detection model for non-colluding wardens, each hop $(u,v)$ is independently monitored by each warden $W_j$.
For warden $W_j$, the DEP for hop $(u,v)$, denoted by $P_{\text{DEP},u,v,j}$, can be calculated from~\eqref{eq:DEP_per_hop_coll} by replacing $\rho_u^{(\text{coll})}$ with the SNR at warden $W_j$ on modality $m_u$, namely $\rho_{u,W_j,m_u}$ in~\eqref{eq:indv_SNR_W_def}.   
The per-hop DEP under non-colluding wardens is then given by
\begin{equation} \label{eq:DEP_hop_nc}
    P_{\text{DEP},u,v}^{(\text{nc})} = \min_{j=1,\ldots,J} P_{\text{DEP},u,v,j}.
\end{equation}
Finally, the overall end-to-end DEP under this model is
\begin{equation} \label{eq:DEP_product_nc}
    P_{\text{DEP},\psi}^{(\text{nc,prod})} = \prod_{(u,v) \in \psi} P_{\text{DEP},u,v}^{(\text{nc})}.
\end{equation}

\section{Joint Route and Resource Optimization} \label{sec:optimization}

In this section, we develop an efficient algorithm to solve the optimization problem in~\eqref{eq:opt_problem}. 
The DEPs for colluding wardens in~\eqref{eq:DEP_exact_coll} and non-colluding wardens in~\eqref{eq:DEP_nc} are determined by the interplay of transmit power, modality selection, and routing.
This joint dependency makes the optimization in~\eqref{eq:opt_problem} fundamentally challenging, as the intricate coupling between physical-layer signal superposition at wardens and network-layer path constraints on the achievable rate renders the joint search space prohibitively large.

To overcome this difficulty, we address the end-to-end covert routing problem through a two-stage decomposition approach.
In the \textit{network-level optimization} stage, we model the network as a weighted directed graph and formulate the end-to-end metric optimization as a shortest-path problem solvable by Dijkstra's algorithm.
This stage defines the link weight $W_{u,v}$ that quantifies the covertness cost of each candidate link as a function of the transmit power $P_u$ and the communication modality $m$.
In the \textit{link-level optimization} stage, we show that the optimal transmit power $\hat{P}_{u,v,m}$ and modality selection $\hat{m}_{u,v}$ for every potential single-hop link $(u,v)$ in the network are uniquely determined by the rate constraint.

\subsection{Network-Level Optimization} \label{subsec:network_opt}

We model the network as a weighted directed graph $G = (V, E, W)$ where $V$ is the set of all nodes, $E$ is the set of feasible links, and $W$ assigns a weight $W_{u,v}$ to each edge for link $(u,v)$.
Dijkstra's algorithm finds the path $\hat{\psi}$ that minimizes the total accumulated weight $\sum_{(u,v) \in \psi} W_{u,v}$ over all $\psi \in \Psi$, which requires the end-to-end cost to decompose as a \emph{sum of per-hop costs}.
In this subsection, we introduce two Dijkstra-compatible link metrics $W_{u,v,m}$ that satisfy this additivity property.
Each link weight $W_{u,v,m}$ is a function of the transmit power $P_u$ and modality~$m$, and the link-level optimization of $P_u$ and $m$ will be provided in Section~\ref{subsec:link_opt}.

\subsubsection{Proposed KL Divergence-Based Metric} \label{subsec:KL_metric}

We derive the proposed link weights from the KL divergence-based lower bounds established in Sections~\ref{subsec:KL_coll} and~\ref{subsec:KL_nc}.

\paragraph{Colluding Case.}
Starting from the KL divergence in~\eqref{eq:KL_coll}, we apply the inequality $\ln(1+x) - \frac{x}{1+x} \leq \frac{x^2}{2}$ for $x \geq 0$ to each term in the sum:
\begin{equation} \label{eq:KL_ub_step1}
    \mathcal{D}^{(\text{coll})}(\mathbb{P}_0 \| \mathbb{P}_1) \leq \frac{L}{2} \sum_{m \in \mathcal{M}_{\psi}} \sum_{j=1}^{J}  \lambda_{m,j}^2.
\end{equation}
Next, since $\lambda_{m,j} \geq 0$, we have $\sum_{j=1}^J \lambda_{m,j}^2 \leq \left(\sum_{j=1}^J \lambda_{m,j}\right)^2$.
As the sum of eigenvalues equals the trace of the whitened SNR matrix, i.e., $\sum_{j} \lambda_{m,j} = \text{tr}(\tilde{\mathbf{R}}_m)$, we obtain
\begin{equation} \label{eq:KL_ub_step2}
    \mathcal{D}^{(\text{coll})}(\mathbb{P}_0 \| \mathbb{P}_1) \!\leq  \frac{L}{2}\!\! \sum_{m \in \mathcal{M}_{\psi}} \!\!\Bigg( \sum_{j=1}^{J}  \lambda_{m,j}  \!\Bigg)^2 \!\!=\! \frac{L}{2}\!\! \sum_{m \in \mathcal{M}_{\psi}} \!\!\! \left( \text{tr}(\tilde{\mathbf{R}}_m) \right)^2.
\end{equation}
By using the linearity of the trace and the definition of $\tilde{\mathbf{R}}_m$ in~\eqref{eq:whitened_SNR}, the trace reduces to
\begin{align} \label{eq:trace_expansion}
    \text{tr}(\tilde{\mathbf{R}}_m) &= \sum_{u \in \mathcal{T}_{\psi}} \mathbb{I}(m_u=m) P_u \|\tilde{\mathbf{h}}_{u,m}\|^2 \nonumber \\
    &= \sum_{u \in \mathcal{T}_{\psi}} \mathbb{I}(m_u=m) \sum_{j=1}^{J} \rho_{u,W_j,m}.
\end{align}
Substituting this back and applying the inequality $\sum_i x_i^2 \le (\sum_i x_i)^2$ for $x_i \geq 0$, we derive an upper bound of $\mathcal{D}^{(\text{coll})}(\mathbb{P}_0 \| \mathbb{P}_1)$ as
\begin{align} \label{eq:KL_coll_bound}
    \mathcal{D}^{(\text{coll})}(\mathbb{P}_0 \| \mathbb{P}_1) &\leq \frac{L}{2} \sum_{m \in \mathcal{M}_{\psi}} \Bigg( \sum_{u \in \mathcal{T}_{\psi}} \mathbb{I}(m_u=m) \sum_{j=1}^{J} \rho_{u,W_j,m} \Bigg)^2 \nonumber \\
    &\leq \frac{L}{2}  \Bigg( \sum_{u \in \mathcal{T}_{\psi}} \sum_{j=1}^{J} \rho_{u,W_j,m_u} \Bigg)^2.
\end{align}

From~\eqref{eq:DEP_LB_coll_def}, the lower bound on the DEP is maximized when the inner sum in~\eqref{eq:KL_coll_bound}, which is additive over hops, is minimized. 
Consequently, the link weight is defined as
\begin{equation} \label{eq:KL_link_weight}
    W_{u,v,m}^{(\text{KL})} = \sum_{j=1}^{J} \rho_{u,W_j,m}.
\end{equation}

\paragraph{Non-Colluding Case.}
Starting from the per-warden KL divergence in~\eqref{eq:KL_nc_j} and applying the bounds $\ln(1+x) - \frac{x}{1+x} \leq \frac{x^2}{2}$ and $\sum_i x_i^2 \leq (\sum_i x_i)^2$, we obtain
\begin{equation} \label{eq:KL_nc_step1}
    \mathcal{D}_{\psi,j} \leq \frac{L}{2} \sum_{m \in \mathcal{M}_{\psi}} \left( \rho_{W_j,m}^{(\text{agg})} \right)^2 \leq \frac{L}{2} \bigg( \sum_{m \in \mathcal{M}_{\psi}} \rho_{W_j,m}^{(\text{agg})} \bigg)^2.
\end{equation}
Expanding each aggregate SNR $\rho_{W_j,m}^{(\text{agg})} = \sum_{u \in \mathcal{T}_{\psi}} \mathbb{I}(m_u = m) \rho_{u,W_j,m}$ and noting that each transmitter~$u$ contributes to exactly one modality, the sum over modalities collapses to a sum over hops:
\begin{equation} \label{eq:KL_nc_bound}
    \mathcal{D}_{\psi,j} \leq \frac{L}{2} \Bigg( \sum_{u \in \mathcal{T}_{\psi}} \rho_{u,W_j,m_u} \Bigg)^2.
\end{equation}

Since the non-colluding DEP is governed by the worst-case warden, the route must minimize $\max_{j} \sum_{u \in \mathcal{T}_{\psi}} \rho_{u,W_j,m_u}$, which is a min-max problem not directly solvable by Dijkstra's algorithm.
Thus, we apply the relaxation $\max_{j} f_j \leq \sum_{j} f_j$ for non-negative $f_j$, giving
\begin{equation} \label{eq:KL_nc_relax}
    \max_{j=1,\ldots,J} \sum_{u \in \mathcal{T}_{\psi}} \rho_{u,W_j,m_u} \leq \sum_{u \in \mathcal{T}_{\psi}} \sum_{j=1}^{J} \rho_{u,W_j,m_u}.
\end{equation}
Finally, the resulting link weight becomes
\begin{equation} \label{eq:KL_link_weight_nc}
    W_{u,v,m}^{(\text{KL})} = \sum_{j=1}^{J} \rho_{u,W_j,m},
\end{equation}
which is identical to that in the colluding case in~\eqref{eq:KL_link_weight}.

\begin{remark} \label{remark:KL_identical}
The proposed KL-based link weights are identical for colluding and non-colluding wardens.
Both reduce to the total SNR leakage $\sum_{j=1}^{J} \rho_{u,W_j,m}$, which is a simple closed-form expression.
This universality is a practical advantage: a single set of link weights serves both warden models, eliminating the need for separate optimization pipelines.
\end{remark}

\subsubsection{Conventional Per-Hop Detection-Based Metric} \label{subsec:prod_metric}

We also consider a conventional metric based on the per-hop detection in Sections~\ref{subsec:prod_DEP_coll} and~\ref{subsec:prod_DEP_nc}, which generalizes the single-warden routing framework of~\cite{Kong:24TIFS} to the multiple-warden case.
The corresponding end-to-end DEPs are defined in~\eqref{eq:DEP_product_coll} and~\eqref{eq:DEP_product_nc}, and the additivity property is obtained by taking the negative logarithm of the product: $-\ln \prod_{(u,v)} P_{\text{DEP},u,v} = \sum_{(u,v)} (-\ln P_{\text{DEP},u,v})$.
The resulting link weights are
\begin{align}
    W_{u,v,m}^{(\text{Prod,coll})} &= -\ln P_{\text{DEP},u,v}^{(\text{coll})}\!\left(\rho_{u}^{(\text{coll})}\right),  \label{eq:prod_link_weight_coll} \\
    W_{u,v,m}^{(\text{Prod,nc})} &= -\ln P_{\text{DEP},u,v}^{(\text{nc})}\!\left(\{\rho_{u,W_j,m}\}_{j=1}^{J}\right), \label{eq:prod_link_weight_nc}
\end{align}
where $P_{\text{DEP},u,v}^{(\text{coll})}$ and $P_{\text{DEP},u,v}^{(\text{nc})}$ are the per-hop DEPs defined in~\eqref{eq:DEP_per_hop_coll} and~\eqref{eq:DEP_hop_nc}, respectively.
Unlike the proposed KL-based weights, the conventional per-hop DEP weights require evaluating the incomplete gamma function for each candidate link, which incurs a higher computational cost as analyzed in Section~\ref{subsec:complexity}.

\subsection{Link-Level Optimization} \label{subsec:link_opt}

We now show that, regardless of which link weight is chosen in Section~\ref{subsec:network_opt}, the optimal transmit power for every link is uniquely determined by the rate constraint.
This result allows the link-level optimization to be performed independently for each link $(u,v)$.

\subsubsection{Rate Monotonicity}

We first show that the rate $R_{u,v}$ of link $(u, v)$ in~\eqref{eq:rate_hop} is a monotonically increasing function of the transmit power $P_{u}$.
Since the SNR $\rho_{u,v,m_u} = \frac{P_u |h_{u,v,m_u}|^2}{N_{0,v} \Omega_{m_u}}$ is linearly proportional to the transmit power $P_u$, it suffices to show that the rate $R_{u,v}$ is strictly increasing in $\rho_{u,v,m_u}$.
The rate expression in~\eqref{eq:rate_hop} can be rewritten as
\begin{equation} \label{eq:rate_wrt_SNR}
    R_{u,v} = \frac{\Omega_{m_u}}{\ln 2} \bigg[ \ln\!\left(1 + \rho_{u,v,m_u}\right) - K \sqrt{1 \!-\! (1 + \rho_{u,v,m_u})^{-2}} ~ \bigg],
\end{equation}
where $K = \frac{Q^{-1}(\epsilon)}{\sqrt{L}} > 0$ since the decoding error probability should satisfy $\epsilon < 0.5$ under standard reliability requirements.
Let $x \triangleq 1 + \rho_{u,v,m_u} \geq 1$ and define $g(x) \triangleq \ln(x) - K\sqrt{1 - x^{-2}}$, which is proportional to $R_{u,v}$.
Differentiating with respect to $x$ yields
\begin{equation}
    g'(x) = \frac{1}{x} - \frac{K}{x^2\sqrt{x^2 - 1}}.
\end{equation}
Thus, $g'(x) > 0$ (strict monotonicity) requires
\begin{equation} \label{eq:condition_mono}
    K < x\sqrt{x^2 - 1},
\end{equation}
while $g(x) > 0$ (positive rate $R_{u,v}$) requires
\begin{equation} \label{eq:condition_pos}
    K < \frac{x \ln(x)}{\sqrt{x^2 - 1}}.
\end{equation}
We show that the positive-rate condition~\eqref{eq:condition_pos} is strictly more restrictive than the monotonicity condition~\eqref{eq:condition_mono} for all $x > 1$, i.e.,
\begin{equation} \label{eq:bound_comparison}
    \frac{x \ln(x)}{\sqrt{x^2 - 1}} < x \sqrt{x^2 - 1} \quad \Longleftrightarrow \quad \ln(x) < x^2 - 1.
\end{equation}
Define $d(x) \triangleq x^2 - 1 - \ln(x)$.
The first and second derivatives are $d'(x) = 2x - 1/x$ and $d''(x) = 2 + 1/x^2$, respectively.
Since $d''(x) \geq 2 > 0$ for all $x \geq 1$, the first derivative $d'(x)$ is strictly increasing on $[1, \infty)$.
Combined with $d'(1) = 1 > 0$, this implies $d'(x) > 0$ for all $x \geq 1$, so $d(x)$ is strictly increasing.
Since $d(1) = 0$, we conclude that $d(x) > 0$ for all $x > 1$, which means that the bound in~\eqref{eq:condition_mono} is always satisfied when the rate is positive.
Therefore, the link rate $R_{u,v}$ is strictly increasing in transmit power $P_u$.

\subsubsection{Link Weight Monotonicity}

Next, we establish that the link weights $W_{u,v,m}^{(\text{KL})}$, $W_{u,v,m}^{(\text{Prod,coll})}$, and $W_{u,v,m}^{(\text{Prod,nc})}$ are strictly increasing in transmit power $P_u$.
Under $\mathcal{H}_0$, the wardens observe only noise, whose distribution is independent of the transmit powers.
Under $\mathcal{H}_1$, the proposed metric $W_{u,v,m}^{(\text{KL})}$ in~\eqref{eq:KL_link_weight} is strictly increasing in $P_u$ since each SNR $\rho_{u,W_j,m_u}$ is linearly proportional to $P_u$. 
Similarly, since the per-hop DEP $P_{\text{DEP},u,v}$ is strictly decreasing in $P_u$, the weights $W_{u,v,m}^{(\text{Prod,coll})}$ in~\eqref{eq:prod_link_weight_coll} and $W_{u,v,m}^{(\text{Prod,nc})}$ in~\eqref{eq:prod_link_weight_nc} are strictly increasing in $P_u$.

\subsubsection{Optimal Power and Modality Selection}

Combining the two monotonicity results, for any given hop $(u,v)$ and modality $m$, the optimal power $\hat{P}_{u,v,m}$ satisfies the rate requirement with equality:
\begin{equation} \label{eq:opt_power}
    R_{u,v}(m, \hat{P}_{u,v,m}) = R_{\text{req}}.
\end{equation}
This reduces the joint optimization over $(m, P_u)$ to a search over modalities alone: for each candidate modality $m$, the minimum feasible power $\hat{P}_{u,v,m}$ is uniquely determined, and the corresponding link weight $W_{u,v,m}(\hat{P}_{u,v,m})$ is computed.
The optimal modality for link $(u,v)$ is then selected as
\begin{equation} \label{eq:opt_modality}
    \hat{m}_{u,v} = \underset{m \in \mathcal{M}}{\arg\min} \; W_{u,v,m}(\hat{P}_{u,v,m}),
\end{equation}
where $W_{u,v,m}(\hat{P}_{u,v,m})$ is evaluated based on the chosen metric from~\eqref{eq:KL_link_weight}, \eqref{eq:prod_link_weight_coll}, or~\eqref{eq:prod_link_weight_nc}.

\begin{algorithm}[ht]
\caption{Proposed Algorithm} \label{alg:covert_routing_multi}
\begin{algorithmic}[1]
\State \textbf{Input:} $V$, $\{h_{u,v,m}\}$, $\{h_{u,W_j,m}\}_{j=1}^{J}$, $\mathcal{M}$, $R_{\text{req}}$, $P_{\text{max}}$, warden model (colluding/non-colluding)
\State \textbf{Initialize:} $W_{u,v} \leftarrow \infty$, $\hat{m}_{u,v} \leftarrow 0$, $\hat{P}_{u,v} \leftarrow 0, \forall u, v$

\State \textbf{Stage 1: Link-Level Optimization}
\For{each link $(u,v) \in V \times V$ such that $u \neq v$}
    \For{each modality $m \in \mathcal{M}$}
        \State Find $\hat{P}_{u,v,m}$ s.t. $R_{u,v}(m, \hat{P}_{u,v,m}) = R_{\text{req}}$
        \If{$\hat{P}_{u,v,m} \le P_{\text{max}}$}
            \State Compute $W_{u,v,m}(\hat{P}_{u,v,m})$ 
            \If{$W_{u,v,m}(\hat{P}_{u,v,m}) < W_{u,v}$}
                \State $\hat{P}_{u,v} \leftarrow \hat{P}_{u,v,m}$, $\hat{m}_{u,v} \leftarrow m$
                \State $W_{u,v} \leftarrow W_{u,v,m}(\hat{P}_{u,v,m})$
            \EndIf
        \EndIf
    \EndFor
\EndFor

\State \textbf{Stage 2: Network-Level Optimization}
\State Run Dijkstra's Algorithm on $G(V, E, W)$ to find the shortest path $\hat{\psi}$
\State Retrieve $\hat{m}_u$ and $\hat{P}_u$ for all active links $(u,v) \in \hat{\psi}$
\If{$\hat{\psi} \neq \emptyset$}
    \State \textbf{Output:} Route $\hat{\psi}$, modalities $\{\hat{m}_u\}$, powers $\{\hat{P}_u\}$
\Else
    \State \textbf{Output:} Infeasible
\EndIf
\end{algorithmic}
\end{algorithm}

\subsection{Algorithm Description} \label{subsec:algorithm}

The proposed algorithm for both warden models operates in two stages.
The first stage performs the link-level power and modality optimization for each link $(u,v)$  via~\eqref{eq:opt_power} and~\eqref{eq:opt_modality}.
The second stage constructs a weighted directed graph $G = (V, E, W)$ using the optimized weights, where the metric is selected from~\eqref{eq:KL_link_weight}, \eqref{eq:prod_link_weight_coll}, or~\eqref{eq:prod_link_weight_nc}, and then applies Dijkstra's algorithm to determine the optimal route $\hat{\psi}$.
The complete procedure is summarized in Algorithm~\ref{alg:covert_routing_multi}.

\subsection{Complexity Analysis} \label{subsec:complexity}

We compare the complexity of three approaches: (i) the optimal exhaustive search method, (ii) the proposed KL-based Dijkstra, and (iii) the conventional per-hop DEP-based Dijkstra.
Let $N = |V|$ denote the number of nodes and $\tau_c > 0$ the target numerical precision for DEP-related computations.
All three approaches share a common per-link subroutine: solving the rate equation~\eqref{eq:opt_power} via bisection over $M$ candidate modalities. 
We define the resulting per-link optimization cost as $C_l \triangleq M \log(1/\tau_b)$, where $\tau_b$ is the bisection precision.

\subsubsection{Optimal Exhaustive Search}

The exhaustive search enumerates all route and modality combinations.
With $\mathcal{O}(N!)$ loop-free paths (worst case) and $M^{N-1}$ modality assignments per route of at most $N{-}1$ hops, the modality is predetermined for each hop, so the per-hop bisection cost reduces to $C_l / M$.
For each candidate, the exact DEP~\eqref{eq:DEP_exact_coll} or~\eqref{eq:DEP_nc} requires eigendecomposing the $J \times J$ matrix $\tilde{\mathbf{R}}_m$ at $\mathcal{O}(MJ^3)$ and evaluating the Gil-Pelaez integral, where each of $\mathcal{O}(\log(1/\tau_c))$ quadrature points costs $\mathcal{O}(MJ)$ to evaluate the characteristic function in~\eqref{eq:CF_coll} or~\eqref{eq:CF_nc}.
The overall complexity is
\begin{equation} \label{eq:complexity_exhaustive}
    \mathcal{O}\!\left( N! \cdot M^{N-1} \left( N C_l / M + M J^3 + M J \log(1 / \tau_c) \right) \right).
\end{equation}

\subsubsection{Proposed KL-Based Dijkstra Algorithm}

The KL-based weight~\eqref{eq:KL_link_weight} sums $J$ precomputed SNRs, costing $\mathcal{O}(J)$ per link.
Over $\mathcal{O}(N^2)$ links, the metric evaluation costs $\mathcal{O}(N^2 M J)$.
Dijkstra's algorithm adds $\mathcal{O}(N^2 \log N)$, giving an overall complexity of
\begin{equation} \label{eq:complexity_proposed}
    \mathcal{O}\!\left(N^2 (C_l + M J + \log N)\right).
\end{equation}

\subsubsection{Conventional Per-Hop DEP-Based Dijkstra Algorithm}

Each per-hop DEP evaluation~\eqref{eq:DEP_per_hop_coll} or~\eqref{eq:DEP_hop_nc} requires computing $\gamma(L, \cdot)$ at $\mathcal{O}(L + \log(1/\tau_c))$ per call.
In the worst case (non-colluding), this is performed for all $J$ wardens, yielding a per-link metric cost of $\mathcal{O}(J(L + \log(1/\tau_c)))$ and an overall complexity of
\begin{equation} \label{eq:complexity_benchmark}
    \mathcal{O}\!\left(N^2 \!\left(C_l + M J \!\left(L + \log(1/ \tau_c )\right) + \log N\right)\right).
\end{equation}

\subsubsection{Comparison}

The proposed and conventional algorithms share the same polynomial route search cost, but the former reduces the per-link metric cost from $\mathcal{O}(J(L + \log(1/\tau_c)))$ to $\mathcal{O}(J)$, yielding a speedup factor of $\mathcal{O}(L + \log(1/\tau_c))$, by replacing gamma function evaluations with SNR summations.
The exhaustive search incurs a combinatorial $\mathcal{O}(N! \cdot M^{N})$ enumeration cost, making it intractable when $N$ or $M$ is not small.

\section{Simulation Results} \label{sec:results}

\begin{figure}[t]
\begin{center} 
\includegraphics[width=2.8in]{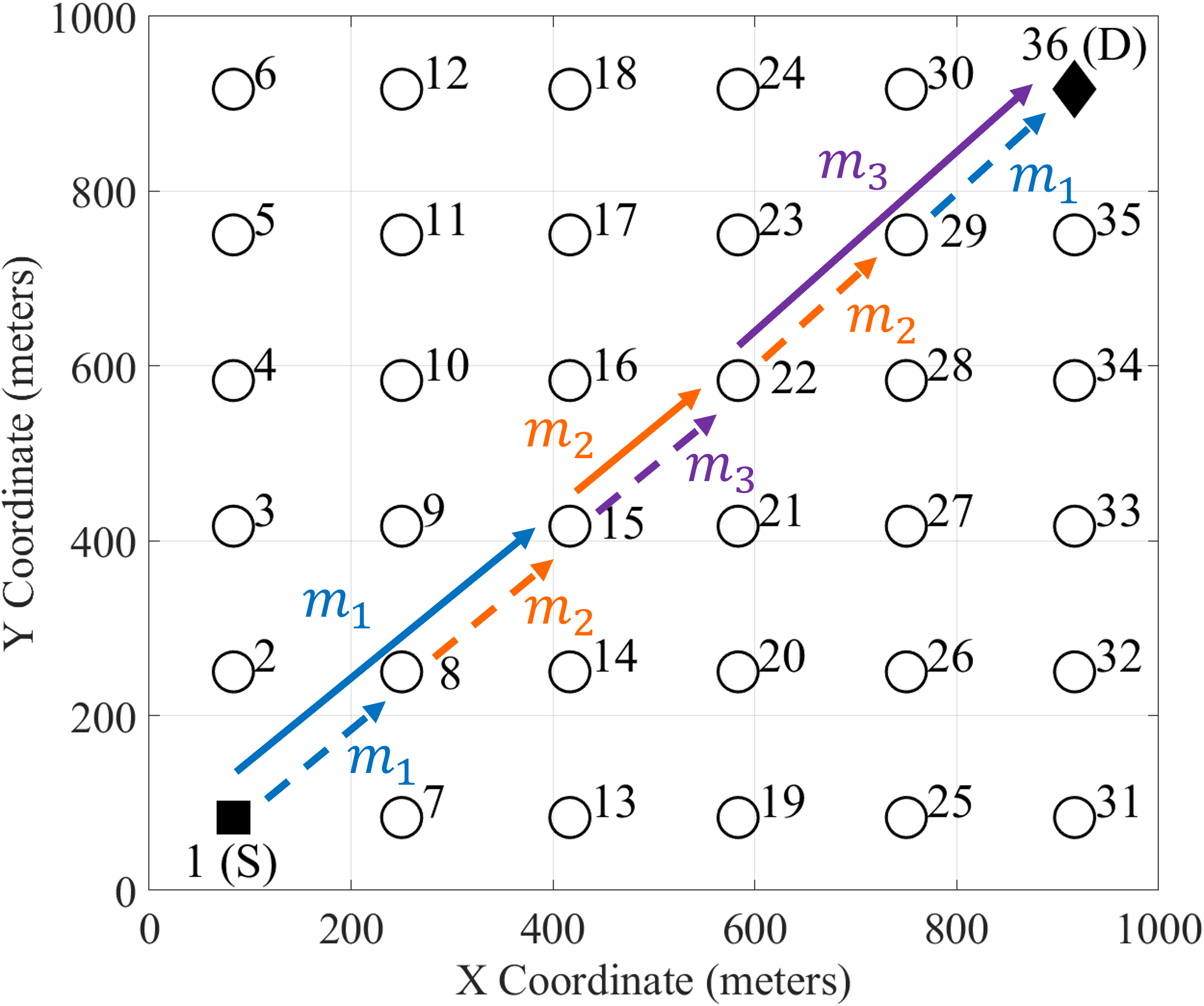}
\end{center} 
\vspace{-5.0mm}
\caption{A network topology with 36 nodes, illustrating two exemplary routes: Route~1 (solid lines) and Route~2 (dashed lines). The individual hops utilizing modalities $m_1$, $m_2$, and $m_3$ are highlighted in blue, orange, and purple, respectively.}
\label{figure:topology}
\end{figure}

In this section, we present numerical simulation results to validate the theoretical analysis developed in Sections~\ref{sec:colluding_analysis} and~\ref{sec:noncolluding_analysis}, and to evaluate the efficiency of the joint optimization framework proposed in Section~\ref{sec:optimization}.
We consider a network environment consisting of 36 nodes as illustrated in Fig.~\ref{figure:topology}, where nodes 1 and 36 serve as the source and destination, respectively, and the remaining 34 nodes act as potential relays for establishing a route. 
We assume that multiple wardens are randomly distributed throughout the network area.
Furthermore, the channel coefficient between any two nodes is given by $h_m = \tilde{h}_m \sqrt{ PL_m(d)}$ where $\tilde{h}_m$ denotes the small-scale Rayleigh fading coefficient.
The large-scale path loss is modeled as $PL_m(d) = -10 \textcolor{red}{n_m} \log_{10}(d) - 20 \log_{10}(4\pi / \lambda_m)$~dB~\cite{Perez:02} where $d$ is the inter-node distance, \textcolor{red}{$n_m$} is a parameter capturing the propagation effects, and $\lambda_m$ is the signal wavelength in meters for modality~$m$. 
Specifically, three communication modalities ($M=3$), $m_1$, $m_2$, and $m_3$, are employed, operating at center frequencies of $600$~MHz, $900$~MHz, and $1200$~MHz, respectively. 
Unless otherwise stated, the network parameters are set as follows: $P_{\text{max}} = 1$~W, $L = 100$, $N_{0,W_j}=N_{0,v}=-120$~dBm/Hz, $\Omega_m = 10$~MHz, and $\epsilon = 0.05$~\cite{Yan:19b}.

\subsection{DEP Analysis}

We obtain both analytical and simulated results by averaging over $10^4$ independent random realizations of the warden locations and inter-node channel fading coefficients.
The evaluation considers the two routes illustrated in Fig.~\ref{figure:topology} assuming a uniform transmit power $P_u = P_{\text{max}}, \forall u$.

\begin{figure}[t]
\begin{center} 
\includegraphics[width=3.2in]{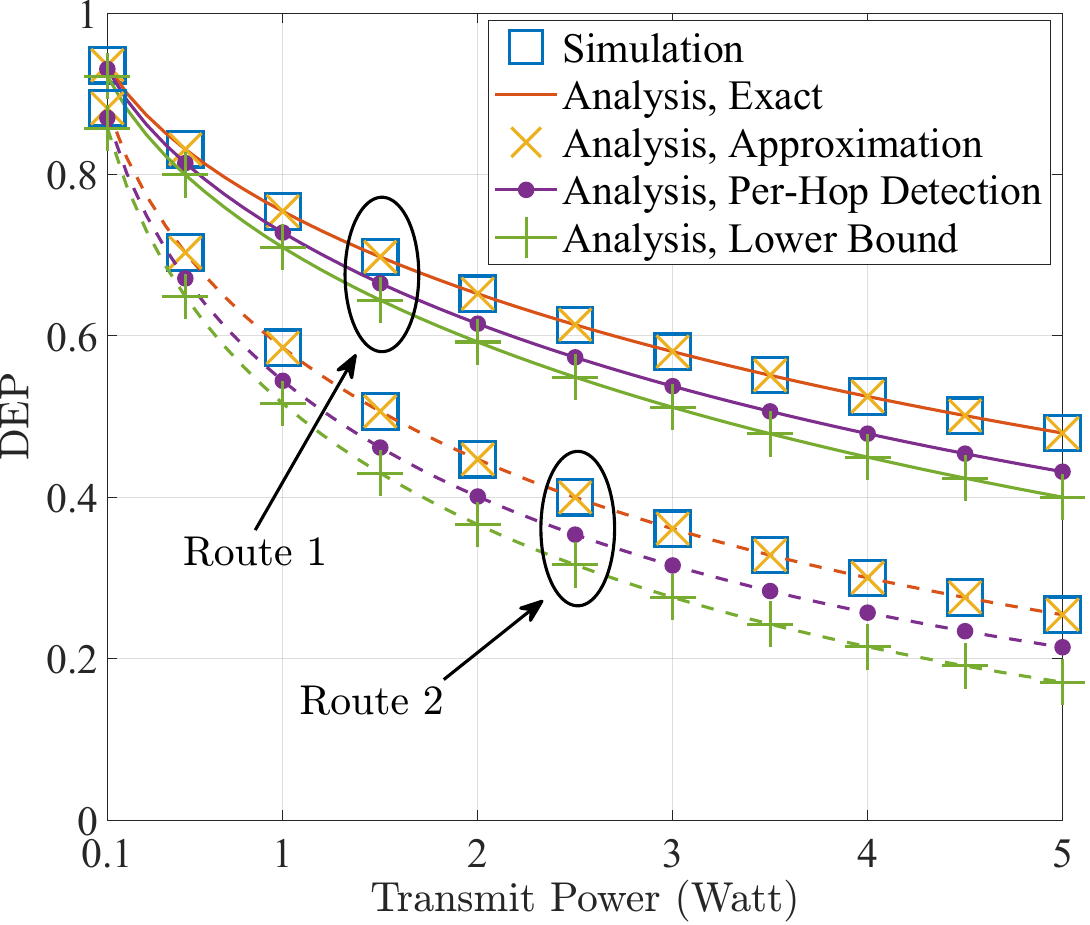}
\end{center} 
\vspace{-5.0mm}
\caption{DEP versus transmit power for the two exemplary routes in the presence of a single warden.}
\label{figure:Analysis_Num_W_1}
\end{figure}

In Fig.~\ref{figure:Analysis_Num_W_1}, we evaluate the DEP of the two routes from Fig.~\ref{figure:topology} against a single warden.
As expected, the DEP monotonically declines as the transmit power increases, driven by the higher aggregate SNR at the warden.
As a result, the 3-hop Route~1 achieves a significantly higher DEP than the 5-hop Route~2, since utilizing fewer simultaneous transmitters minimizes the total energy leakage.
From an analytical perspective, we observe that the exact expressions perfectly match the Monte Carlo simulation results.
Moreover, the proposed moment-matching-based approximation is remarkably tight across the entire power regime.
In contrast, the conventional product of per-hop DEPs consistently underestimates the exact DEP, while the KL divergence-based lower bound proves relatively loose and overly conservative.

\begin{figure}[t]
\begin{center} 
\includegraphics[width=3.2in]{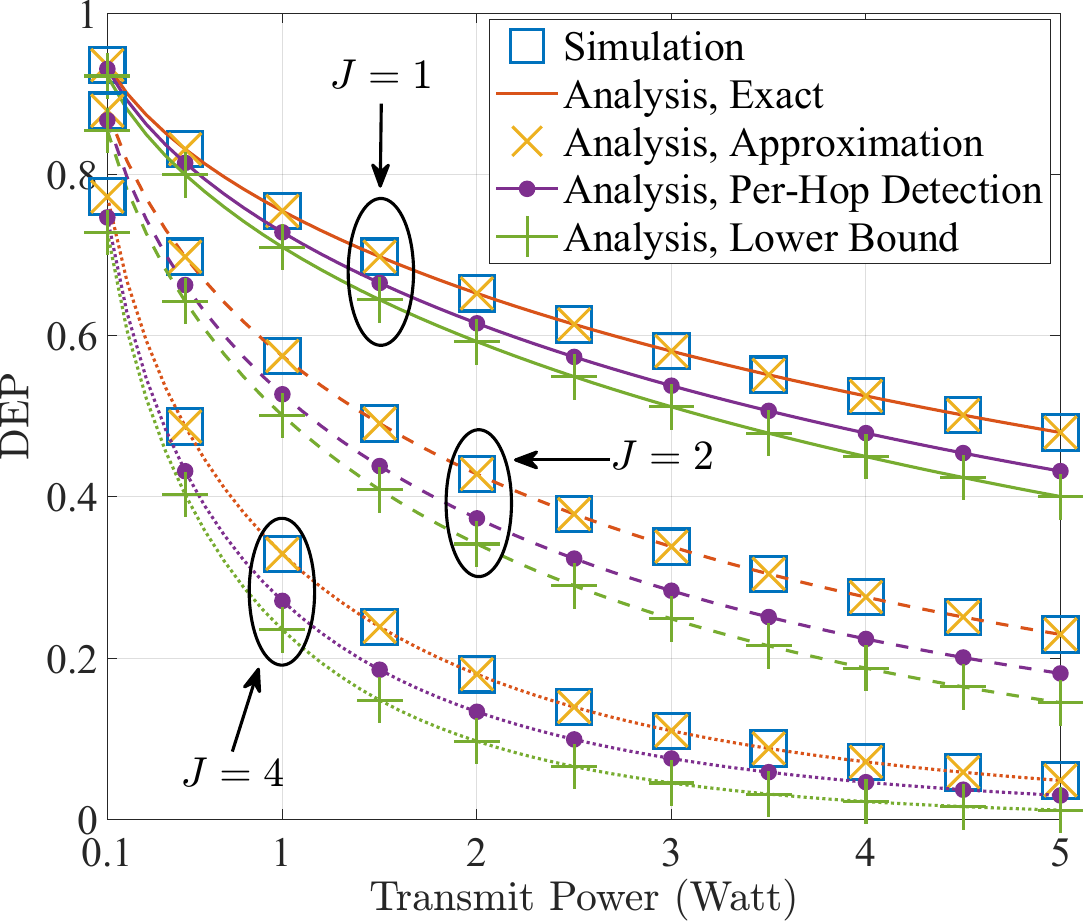}
\end{center} 
\vspace{-5.0mm}
\caption{DEP versus transmit power for Route~1 against varying numbers of colluding wardens $J$.}
\label{figure:Analysis_Colluding}
\end{figure}

Fig.~\ref{figure:Analysis_Colluding} demonstrates the DEP of Route~1 against varying numbers of colluding wardens~$J$.
As anticipated, the DEP consistently decreases as $J$ increases and declines more rapidly with transmit power for larger values of~$J$.
This severe degradation occurs because the central fusion center effectively accumulates spatially distributed signal energies to enhance joint detection.
Additionally, the steepest DEP drops occur in the low-power regime, highlighting the critical need for precise power control at low SNRs.
Finally, the results confirm that the proposed moment-matching approximation remains in close agreement with the exact DEP, while both the product of per-hop DEPs and the KL divergence lower bound underestimate the actual DEP.

\begin{figure}[t]
\begin{center} 
\includegraphics[width=3.2in]{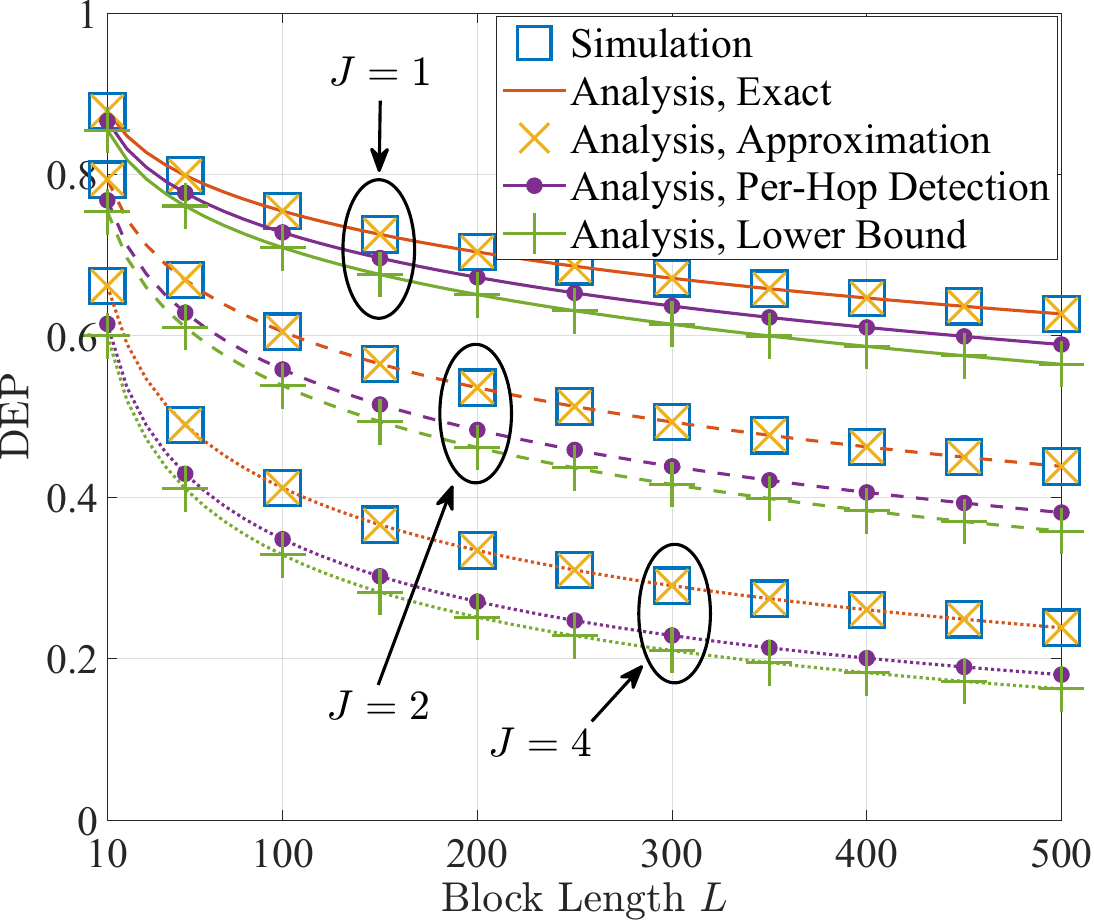}
\end{center} 
\vspace{-5.0mm}
\caption{DEP versus blocklength $L$ for Route~1 against varying numbers of non-colluding wardens $J$.}
\label{figure:Analysis_Noncolluding}
\end{figure}

In Fig.~\ref{figure:Analysis_Noncolluding}, we present the DEP of Route~1 as a function of blocklength~$L$ under the non-colluding warden model.
The DEP steadily decreases as $L$ grows, since longer observation windows allow the wardens to accumulate more signal energy. 
Similarly, the overall DEP declines as the number of wardens~$J$ increases, with the steepest degradation occurring at short blocklengths.
Notably, the DEP curves for multiple wardens closely mirror those of a single warden with a mere downward shift.
This behavior indicates that, without coherent signal integration, system covertness is dominated by the single warden experiencing the strongest physical channel.
Lastly, we observe that both the exact and approximate analytical results track the simulated results very well across all evaluated blocklengths.

\begin{figure}[t]
\begin{center} 
\includegraphics[width=3.2in]{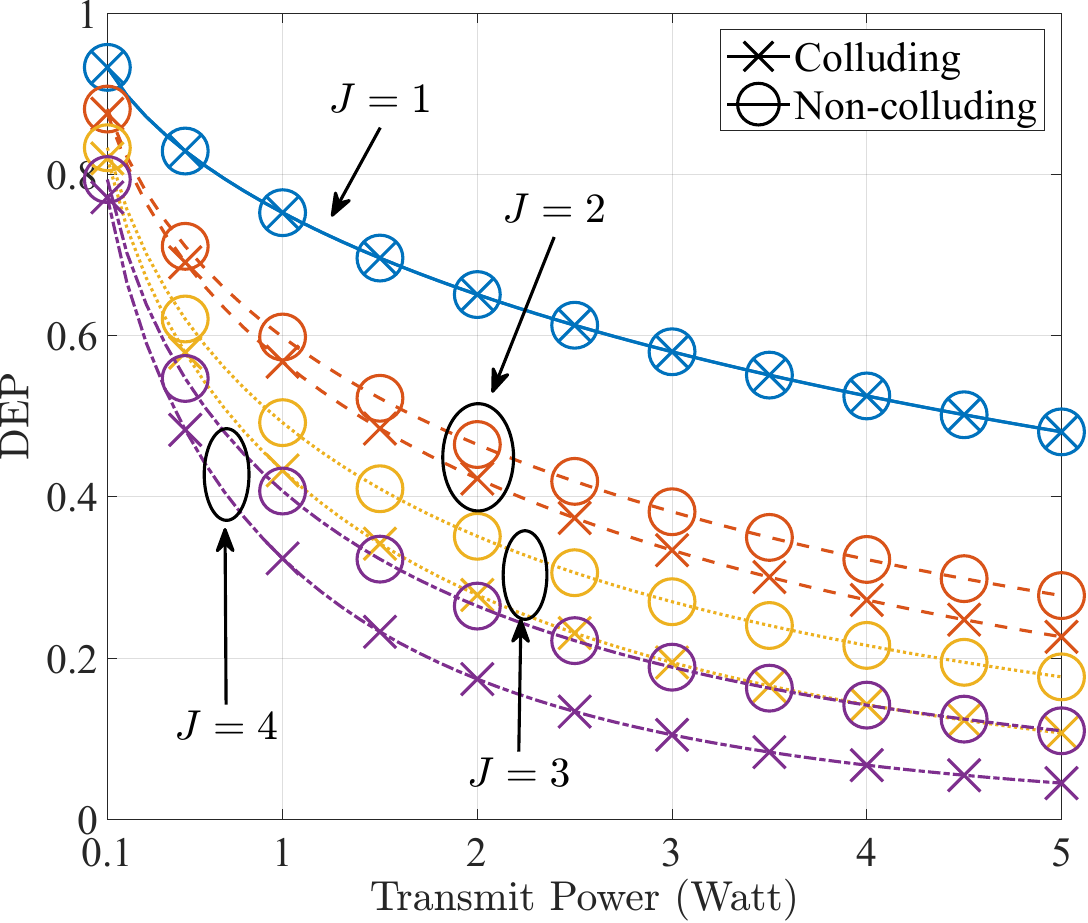}
\end{center} 
\vspace{-5.0mm}
\caption{DEP comparison between the colluding and non-colluding warden models for Route~1 across various numbers of wardens $J$.}
\label{figure:Analysis_CollvsNonc}
\end{figure}

Fig.~\ref{figure:Analysis_CollvsNonc} compares the exact DEP of Route~1 under the colluding and non-colluding warden models across various numbers of wardens~$J$.
As expected, for any $J > 1$, the colluding wardens achieve a markedly lower DEP than their non-colluding counterparts.
We also observe that the performance gap between the colluding and non-colluding configurations widens as $J$ increases.
This superior detection capability arises because information sharing and coherent signal integration at the fusion center inherently outperform independent per-warden decision-making.
Therefore, to maintain a target DEP, the allowable transmit power must be significantly reduced as the number of wardens grows, particularly against colluding surveillance networks.

\subsection{Joint Route and Resource Optimization}

In this subsection, we present the DEP performance achieved by the joint route and resource optimization strategies detailed in Section~\ref{sec:optimization}.
The simulated results are obtained by averaging over $10^3$ independent random realizations of the warden locations and inter-node channel coefficients.

\begin{figure}[t]
\begin{center} 
\includegraphics[width=3.2in]{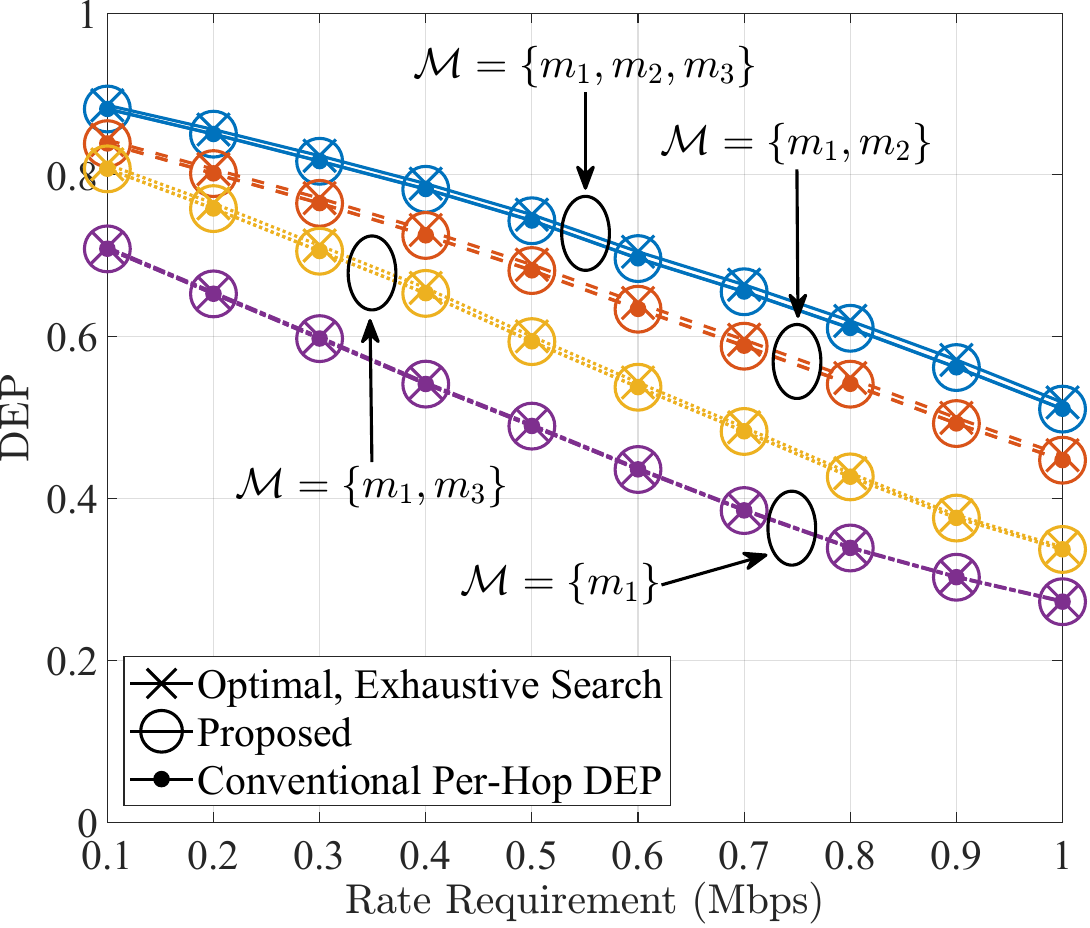}
\end{center} 
\vspace{-5.0mm}
\caption{Optimized DEP versus rate requirement $R_{\text{req}}$ for various available modality sets under the colluding warden model ($J=2$).}
\label{figure:Optimization_Colluding}
\end{figure}

In Fig.~\ref{figure:Optimization_Colluding}, we evaluate the DEP of the proposed algorithm against the optimal exhaustive search assuming two colluding wardens.
We consider various modality sets to highlight the benefits of spectral diversity. 
A key observation is that the proposed framework achieves near-optimal performance across all rate requirements while requiring significantly reduced computational complexity compared to the exhaustive search.
Moreover, expanding the available modalities from a single band to the full set $\{m_1, m_2, m_3\}$ substantially enhances the DEP, as additional spectral degrees of freedom allow the algorithm to bypass wardens via modalities with more favorable propagation characteristics. 
It is also worth noting that while the conventional per-hop DEP-based routing provides nearly identical performance to the proposed KL-based metric, it incurs a higher computational overhead due to the repeated evaluation of incomplete gamma functions.
These results validate that the proposed technique effectively exploits multi-modal diversity to sustain robust covertness with high efficiency.

\begin{figure}[t]
\begin{center} 
\includegraphics[width=3.2in]{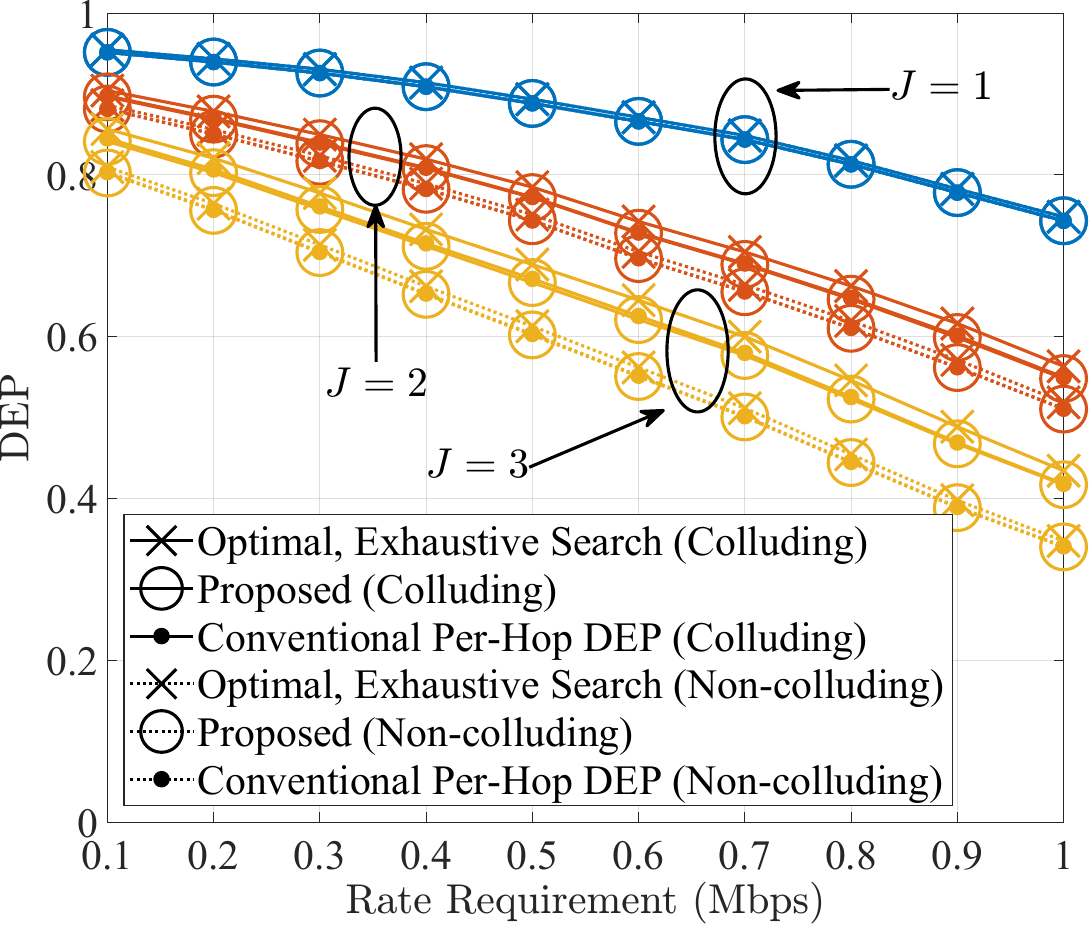}
\end{center} 
\vspace{-5.0mm}
\caption{Optimized DEP versus rate requirement $R_{\text{req}}$ under colluding and non-colluding warden models for varying numbers of wardens $J$.}
\label{figure:Optimization_CollvsNonc}
\end{figure}

Fig.~\ref{figure:Optimization_CollvsNonc} demonstrates the optimized DEP under both colluding and non-colluding warden models for various numbers of wardens~$J$. 
First, the proposed algorithm exhibits near-optimal performance under both models across all rate requirements. 
We observe that as the rate requirement $R_{\text{req}}$ increases, the DEP for both models deteriorates due to the higher transmit power necessary to meet stringent throughput constraints.
As anticipated, the optimized DEP for the non-colluding case is consistently higher than that of the colluding case, as the lack of centralized information sharing significantly limits the wardens' joint detection capability.
Additionally, the performance gap between the two models widens as the number of wardens~$J$ grows, highlighting that the detection gain from collusion increases for larger values of~$J$.

\begin{figure}[t]
	\centering
	\subfigure[Average number of hops of the optimized routes.]{  \hspace{-5mm}
		\includegraphics[width=.45\textwidth]{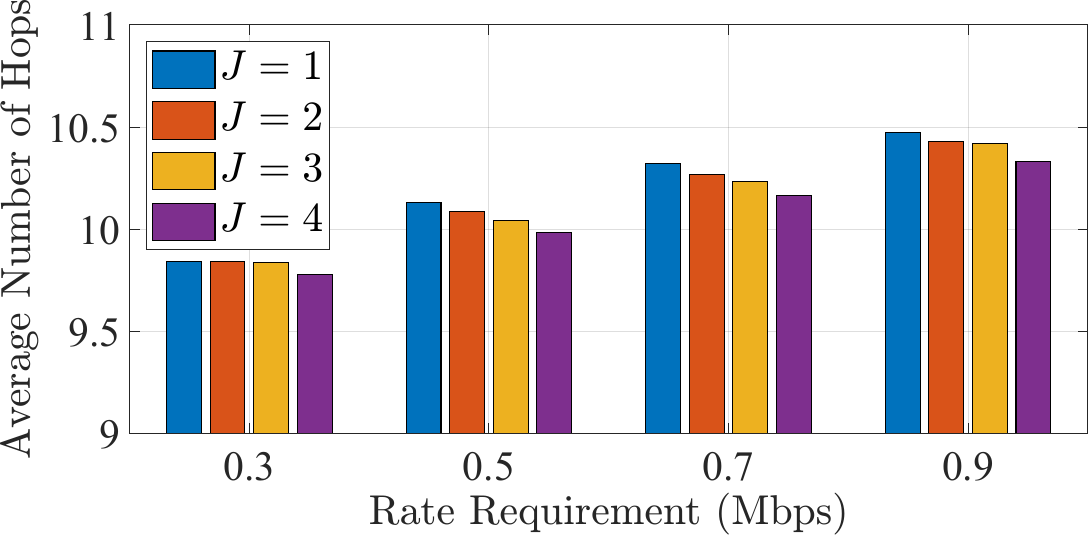} 
	}
	\vfill
	\subfigure[Modality selection probabilities for $\{m_1, m_2, m_3\}$.]{ \hspace{-5mm}
		\includegraphics[width=.45\textwidth]{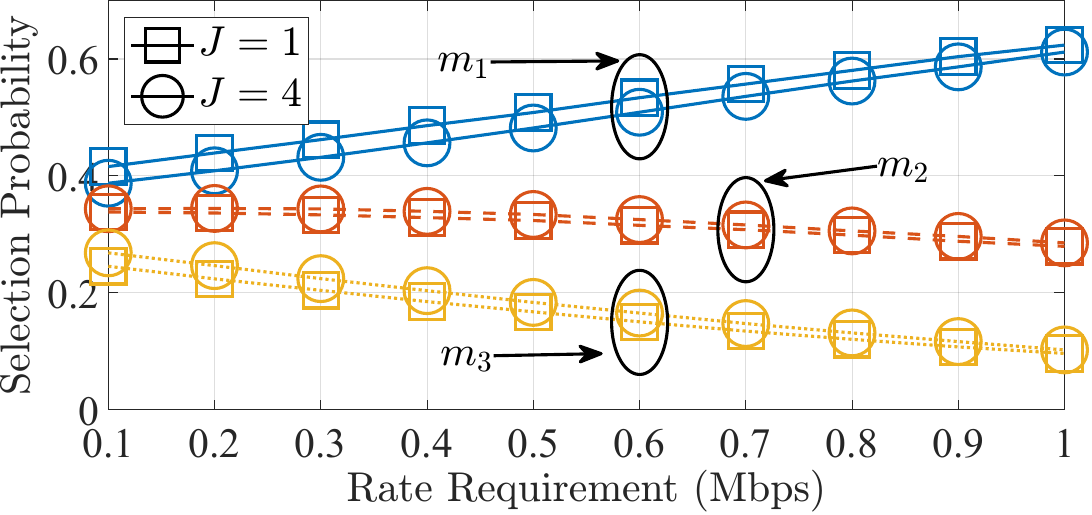}
	}
    \vspace{-2mm}
	\caption{Average number of hops and modality selection probabilities versus rate requirement $R_{\text{req}}$ under the colluding warden model.}
	\label{fig:Figure_Optimization_Hops_Mods}
\end{figure}

In Fig.~\ref{fig:Figure_Optimization_Hops_Mods}, we examine the average number of hops of the optimized routes and the modality selection probabilities under the colluding warden model.
As the rate requirement $R_{\text{req}}$ increases, the average number of hops rises, indicating that short-distance multi-hop transmissions are desirable to satisfy stringent rate constraints. 
Conversely, the number of hops decreases as the number of wardens~$J$ increases, suggesting that reducing the total number of transmission events minimizes the aggregate signal exposure and thereby improves covertness.
The selection probabilities for the three modalities are comparable at low~$R_{\text{req}}$, revealing highly adaptive behavior.
However, as $R_{\text{req}}$ becomes more stringent, the selection probability of~$m_1$ surges since higher rates are more effectively attained by leveraging the lower operating frequency of~$m_1$, which exhibits higher channel gains relative to the other modalities.
Finally, as~$J$ increases, the selection probabilities become slightly more distributed.
This shift suggests an incremental exploitation of spectral diversity to circumvent colluding wardens, though this effect remains secondary to the impact of the rate requirement.

\section{Conclusion and Future Work} \label{sec:conclusion}

In this paper, we presented a comprehensive study on the analysis and optimization of multi-modal routing for covert communications against both colluding and non-colluding adversarial models.
We first investigated the optimal detectors for each scenario and derived exact DEP expressions.
Subsequently, we established moment-matching-based closed-form approximations, theoretical lower bounds, and DEP formulations under a per-hop detection framework.
Furthermore, we developed a joint route and resource optimization algorithm that decouples link-level resource allocation from network-level path selection by leveraging a novel low-complexity routing metric.
Through extensive numerical simulations, we validated our theoretical derivations and demonstrated that the proposed strategy exhibits near-optimal performance while requiring significantly reduced computational complexity.

Future research could build upon this work by addressing uncertainty regarding the wardens' channel state information, which would necessitate more robust or uncertainty-aware resource allocation schemes.
Additionally, to capture temporal dynamics, integrating queuing theory into the network design could reveal critical latency-covertness trade-offs.
Finally, transitioning toward decentralized architectures capable of managing multiple concurrent flows remains an open challenge, where distributed learning-based or game-theoretic approaches could offer scalable solutions for large-scale covert networks.

\bibliographystyle{ieeetr}
\bibliography{bibliography.bib}

\end{document}